\documentclass[
  journal=pasa,
  manuscript=research-paper, 
  year=2024,
  volume=40,
]{cup-journal}

\usepackage{microtype,siunitx,booktabs,amssymb,amsmath}
\usepackage{gensymb}
\usepackage{bm}
\usepackage{aas-macros}
    \title{BRAiSE: synthetic polarisation in RMHD AGN jet simulations}

\author{L. Jerrim}
\affiliation{School of Natural Sciences, Private Bag 37, University of Tasmania, Hobart, 7001 TAS, Australia}
\email[L. Jerrim]{larissa.jerrim@utas.edu.au}

\author{S. Shabala}
\affiliation{School of Natural Sciences, Private Bag 37, University of Tasmania, Hobart, 7001 TAS, Australia}
\alsoaffiliation{ARC Centre of Excellence for All Sky Astrophysics in 3 Dimensions (ASTRO 3D), Australia}

\author{P. Yates-Jones}
\affiliation{School of Natural Sciences, Private Bag 37, University of Tasmania, Hobart, 7001 TAS, Australia}
\alsoaffiliation{ARC Centre of Excellence for All Sky Astrophysics in 3 Dimensions (ASTRO 3D), Australia}

\author{M. Krause}
\affiliation{Centre for Astrophysics Research, University of Hertfordshire, College Lane, Hatfield, Herts AL10 9AB, UK}

\author{R. J. Turner}
\affiliation{School of Natural Sciences, Private Bag 37, University of Tasmania, Hobart, 7001 TAS, Australia}

\author{G. Stewart}
\affiliation{School of Natural Sciences, Private Bag 37, University of Tasmania, Hobart, 7001 TAS, Australia}

\author{C. Power}
\affiliation{International Centre for Radio Astronomy Research, University of Western Australia, 35 Stirling Highway, Crawley, WA 6009, Australia}
\alsoaffiliation{ARC Centre of Excellence for All Sky Astrophysics in 3 Dimensions (ASTRO 3D), Australia}

\doi{10.1017/pasa.2020.32}

\received {23 Apr 2025}
\revised  {24 Jun 2025}
\accepted {19 Sep 2025}
\published{26 Sep 2025}


\begin{document}

\begin{abstract}

We present a new method to calculate the polarised synchrotron emission of radio AGN sources using magnetic field information from 3-dimensional relativistic magnetohydrodynamical (RMHD) simulations. Like its predecessor, which uses pressure as a proxy for the magnetic field, this method tracks the spatially resolved adiabatic and radiative loss processes using the method adapted from the Radio AGN in Semi-analytic Environments formalism. Lagrangian tracer particles in RMHD simulations carried out using the PLUTO code are used to track the fluid quantities of each `ensemble of electrons' through time to calculate the radio emissivity \textit{ex situ}. By using the magnetic field directly from simulations, the full set of linear Stokes parameters I, Q, and U can be calculated to study the synthetic radio polarisation of radio AGN sources. We apply this method to a suite of RMHD simulations to study their polarisation properties. The turbulent magnetic field present in radio lobes influences the emission, causing a complex clumpy structure that is visible at high resolution. Our synthetic polarisation properties are consistent with observations; we find that the fractional polarisation is highest (approximately $50$ percent) at the lobe edges. We show that for the same source, the integrated and mean fractional polarisation depends on viewing angle to the source. At oblique viewing angles the behaviour of the integrated and mean fractional polarisation over time depends on the morphology of the jet cocoon. Using Faraday rotation measures, we reproduce known depolarisation effects such as the Laing-Garrington depolarisation asymmetry in jets angled to the line of sight. We show that the hotspots and hence the Fanaroff-Riley classification become less clear with our new, more accurate method.

\end{abstract}

\section{Introduction}

Active Galactic Nuclei (AGN) jets and their associated radio lobes are extragalactic sources of radio synchrotron emission. Since AGN jets provide an important source of feedback that regulates galaxy evolution \citep{mcnamara_heating_2007, somerville_physical_2015}, understanding the emission of their radio lobes is critical for interpreting the physical effect that AGN jets have on their environments. The synchrotron emission comes from high-energy electrons in AGN jets that are accelerated at shocks \citep[e.g.][]{burbidge_synchrotron_1956}. These electrons undergo energy losses due to synchrotron, adiabatic, and inverse-Compton (IC) processes. The energy loss processes are dependent upon the dynamics of the jet and its cocoon. In particular, the magnitude and spatial distribution of synchrotron emission depends on the magnetic fields present in AGN jets and their cocoons.

Radio synchrotron emission from AGN jets is highly linearly polarised, with a maximum fractional polarisation of around $70\%$ \citep{pacholczyk_radio_1970}. Simulation studies of polarised radio emission from AGN jet-inflated lobes have been conducted over the past few decades \citep[e.g.][]{matthews_models_1990-1, matthews_models_1990, tregillis_synthetic_2004, huarte-espinosa_3d_2011, hardcastle_numerical_2014, english_numerical_2016, meenakshi_polarization_2023, stimpson_numerical_2023}. In this work, we study the polarisation properties of kpc-scale AGN jets and their associated radio lobes using 3-dimensional relativistic magnetohydrodynamical simulations.

Previous studies of the Stokes parameters in simulations of kpc-scale AGN jets by \citet{huarte-espinosa_3d_2011}, \citet{hardcastle_numerical_2014}, and \citet{english_numerical_2016} calculate the synchrotron emission based on numerical grid quantities; this is the same approach as originally outlined by \citet{matthews_models_1990-1, matthews_models_1990}, though the latter authors derived the magnetic field from simplifying assumptions using purely hydrodynamic simulations. The Stokes emissivities due to synchrotron emission are calculated in a grid-based manner based upon the magnetic field components, which are then integrated along each line of sight to create a 2-dimensional synthetic observation. This method does not account for an evolving electron energy distribution (e.g. one experiencing adiabatic and IC losses) but is a useful first step into understanding how the magnetic structure in the radio lobes influences the emission.

\citet{meenakshi_polarization_2023} have published the most advanced RMHD simulations of polarised radio emission in AGN jets and lobes to date. This method uses a hybrid grid and particle method by \citet{vaidya_particle_2018} to track the evolution of each particle as it is advected by the fluid on the simulation grid. The energy distribution of the particles is followed \textit{in-situ} by solving the relativistic cosmic ray transport equation. The synchrotron emissivity and full Stokes parameters are calculated within the simulations, and the surface brightness images are then ray-traced in post-processing. The authors study radio lobes of intermediate sizes of up to approximately $30$ kpc in total length. This continues a tradition of \textit{in-situ} treatment of electrons in AGN jet simulations \citep[e.g.][]{jones_simulating_1999, tregillis_synthetic_2004, jones_efficient_2005, nolting_observations_2022}.

\citet{kramer_3d_2024} use a similar method in their RMHD simulations of parsec-scale AGN jets. The authors use Lagrangian particles in their simulations and follow the energy distribution of the particles in a similar manner to \citet{vaidya_particle_2018}, however, they post-process their simulated data (including calculating the polarised emission) using the RADMC-3D code \citep{macdonald_electrons_2021}. The RADMC-3D code calculates the full radiative transfer of the Stokes' parameters as based on the methods of \citet{jones_transfer_1977, jones_transfer_1977b}. We do not consider the full radiative transfer of the Stokes' parameters in this work. This hybrid method incorporates a mixture of \textit{in-situ} and \textit{ex-situ} treatments (\textit{in-situ} particles/energy distribution, \textit{ex-situ} emissivity), in contrast to a fully post-processed method \citep[e.g.][]{fromm_spectral_2016} or fully \textit{in-situ} method \citep[e.g.][]{meenakshi_polarization_2023}.

In this paper we introduce a new method of calculating polarised radio emission, building on the PRAiSE (Particles + Radio AGN in Semi-analytic Environments) framework \citep{yates-jones_praise_2022, turner_raise_2018_II}. The PRAiSE method is also a hybrid approach, using Lagrangian particles that are advected by the fluid in a grid-based simulation to evolve the particle energy distribution and calculate synchrotron emissivity \textit{ex-situ}. These particles record the history of the local fluid values (pressure, density, velocity, magnetic field) encountered along their trajectory and the time since the particle last passed through a shock \citep[using three thresholds, see][for more details]{yates-jones_praise_2022}. The particle energy distributions and the associated synchrotron emission are calculated in post-processing, rather than within the simulation, and account for (re-)acceleration due to strong shocks, and subsequent adiabatic, synchrotron, and IC losses. The evolution of particle energy is found by interpolating the emitting particle Lorentz factor backwards in time to the last acceleration event experienced by that particle. In contrast to the methods used by \citet{vaidya_particle_2018}  and \citet{kramer_3d_2024}, our method allows for more flexibility in our parameter choice (e.g. injection index, redshift) as the computationally efficient particle energy distribution calculation is performed in post-processing, rather than \textit{in-situ}.

Previously, the PRAiSE method has made the assumption that the magnetic field is proportional to the pressure in AGN jets and their lobes. In this paper, we directly use the magnetic field from relativistic magnetohydrodynamic simulations to calculate the radio emission, which is implemented as an extension to the PRAiSE code, which we call BRAiSE (B field + RAiSE). Since the magnetic fields in radio lobes are turbulent, the structure of the magnetic field will influence the structures seen in the radio emission \citep{huarte-espinosa_3d_2011, hardcastle_numerical_2014}. This is seen in observations, where the intrinsic magnetic field direction follows synchrotron filaments in radio lobes \citep[e.g.][]{taylor_vla_1990, eilek_magnetic_2002, sebokolodi_wideband_2020}. \\

We use BRAiSE to study depolarisation through calculating polarisation position angles  \citep[e.g.][]{macdonald_electrons_2021, meenakshi_polarization_2023} and Faraday depth \citep[e.g.][]{jerrim_faraday_2024} for the radio synchrotron emission. AGN radio lobes are depolarised through Faraday rotation as the radio emission travels through a magnetoionised medium to the observer \citep{burn_depolarization_1966}. Faraday rotation rotates the polarisation angle of the radio emission in the following way \citep{burn_depolarization_1966}:

\begin{equation}
\label{eqn:chi}
    \chi = \chi_0 + \phi \lambda^2.
\end{equation}

Here, $\chi_0$ is the intrinsic polarisation angle, $\phi$ is the Faraday depth and $\lambda$ is the wavelength of the emission. The Faraday depth quantifies the amount of Faraday rotation and depends on the magnetic field strength, environment density, and line of sight to the source \citep{carilli_cluster_2002}:

\begin{equation}
\label{eqn:RM_BRAiSE}
    \phi = 812 \int^l_0 n_e \bm{B} \cdot \mathrm{d}\bm{l} \: \: \mathrm{rad/m^{2}}.
\end{equation}

Here, the thermal electron number density $n_e$ is in cm$^{-3}$, the magnetic field strength $\bm{B}$ is in $\mu$G, and the path length $\mathrm{d}\bm{l}$ is in kpc. As described in \citet{jerrim_faraday_2024}, we can calculate the Faraday depth from our fluid variables on the simulation grid. In this work, we apply these rotation measures to our generated synchrotron emission to directly study depolarisation by Faraday rotation in our synthetic polarised radio emission.

Depolarisation due to differential Faraday rotation can occur due to the three-dimensional nature of the radio lobe; emission from different depths throughout the lobe will have different amounts of Faraday rotation, which will randomise the polarisation angle \citep{burn_depolarization_1966, sokoloff_depolarization_1998}. However, the radio emission is also depolarised through the observing process. Within a telescope observing beam, many lines of sight to the source will be averaged over: if the polarisation of the emission or the Faraday rotation measure structure is not uniform across the beam, this averaging will reduce the total fractional polarisation \citep{gabuzda_polarization_2020}. This observational effect is known as beam depolarisation.

The paper is structured as follows. In Section \ref{section:braise-eqns}, we present the differences between the equations for when we use pressure (PRAiSE) or magnetic field (BRAiSE) to calculate the radio emission. We validate our results in Section \ref{section:validation} by discussing the jet dynamics (Section \ref{section:dynamics_BRAiSE}) and then the impact of the dynamics and the emission method on the surface brightness (Section \ref{section:radio_BRAiSE}). The synthetic polarised radio images are discussed in Section \ref{section:polarisation_BRAiSE}, including an examination of depolarisation due to Faraday rotation (Section \ref{section:RM}). We conclude with a summary of our findings in Section \ref{section:conclusions_BRAiSE}.

\section{BRAiSE implementation}
\label{section:braise-eqns}

\subsection{Synchrotron emission}
\label{section:synch}

Our synthetic radio synchrotron emission code (PRAiSE) uses the pressure of the Lagrangian tracer particles in our PLUTO simulations, mapping this to the magnetic field energy density for the synchrotron calculation. For the full details of the PRAiSE calculation, we refer the reader to \citet{yates-jones_praise_2022} and \citet{turner_raise_2018_II}. In this work, we introduce the BRAiSE extension to the radio emission code, which calculates the radio emissivities directly from the magnetic field strength of the Lagrangian tracer particles in our simulations. We also include a framework to calculate the full Stokes parameters from our simulations. Removing the assumption of mapping pressure to magnetic energy density changes the particle rest-frame emissivity per unit volume and per unit solid angle at frequency $\nu$ from \citep[Equation 5 of][]{yates-jones_praise_2022}:

\begin{align}
\label{eqn:praise}
\begin{split}
    J_{\nu} = &\frac{\kappa(s)}{4 \pi} \left( \frac{e^2}{2} \right)^{\frac{5 + s}{4}} \left( \frac{\mu_0}{(\Gamma_c - 1)} \right)^{\frac{5 + s}{4}} \left(\frac{3}{\pi}\right)^{\frac{s}{2}} \frac{1}{m_e^{\frac{3 + s}{2}}  c (s+1)} \nu^\frac{1 - s}{2} \\
        & \times \frac{\eta^{\frac{1+s}{4}}}{(\eta + 1)^{\frac{5+s}{4}}}  p(t)^{\frac{5 + s}{4}} \left( \frac{p(t_{\rm acc})}{p(t)} \right)^{1 - \frac{4}{3 \Gamma_c}} \left(\frac{\gamma_{\rm acc}}{\gamma}\right)^{2-s} \\
        & \times \left[\frac{\gamma_{\rm max}^{2-s} - \gamma_{\rm min}^{2-s}}{2 - s} - \frac{\gamma_{\rm max}^{1-s} - \gamma_{\rm min}^{1-s}}{1 - s} \right]^{-1}, \\
\end{split}
\end{align}

to:

\begin{align}
\label{eqn:braise}
    \begin{split}
        J_{\nu} = &\frac{\kappa(s)}{2 \pi} \left( \frac{e^2}{4} \right)^{\frac{5 + s}{4}} \mu_0 \left(\frac{3}{\pi}\right)^\frac{s}{2}  \frac{1}{m_e^{\frac{3+s}{2}} c (s+1)} \nu^\frac{1 - s}{2} \\
        & \times B^{\frac{1+s}{2}} u_e(t_{\rm acc}) \left( \frac{p'(t_{\rm acc})}{p'(t)} \right)^{- \frac{4}{3 \Gamma_c}} \left(\frac{\gamma_{\rm acc}}{\gamma}\right)^{2-s} \\
        & \times \left[\frac{\gamma_{\rm max}^{2-s} - \gamma_{\rm min}^{2-s}}{2 - s} - \frac{\gamma_{\rm max}^{1-s} - \gamma_{\rm min}^{1-s}}{1 - s} \right]^{-1}. \\
    \end{split}
\end{align}

In these equations, $\Gamma_c$ is the cocoon adiabatic index, $s$ is the electron energy power law exponent, $\gamma_{\rm min}$ and $\gamma_{\rm max}$ are the accelerated particle Lorentz limits, and the particle Lorentz factors $\gamma$ and $\gamma_{\rm acc}$ are calculated at the current time and time of particle shock acceleration respectively.

We also have $p(t)$ and $p(t_{acc})$ as the local particle pressures at the current time and time of particle acceleration respectively. In Equation \ref{eqn:praise}, $p$ refers to the total pressure, which includes thermal, non-thermal, and magnetic components. However, in our numerical simulation code, the pressure variable only accounts for non-magnetic pressure. Since in PRAiSE we assume that the magnetic field scales with thermal pressure, the ratio $p(t_{\rm acc})/p(t)$ will remain unchanged whether the non-magnetic or total pressure is used. This ratio comes from adiabatic expansion, which is driven by thermal pressure, so we do not include the magnetic pressure in BRAiSE; we denote the non-magnetic pressure as $p'(t)$. In BRAiSE, we therefore calculate the electron energy density as: 

\begin{align}
\label{eqn:ue}
    \begin{split}
        u_e(t_{\rm acc}) &= \frac{p'(t_{\rm acc})}{(\Gamma_c - 1)}.
    \end{split}
\end{align}

We note that the emissivity equations appear to differ in the power of the adiabatic expansion factor, however, the equations reduce to the same result; the difference is reflected in the differing scaling constants between the two approaches. The constant $\kappa(s)$ is given by \citet{longair_high_2011} to be:

\begin{equation}
    \kappa(s) = \frac{\Gamma(\frac{s}{4} + \frac{19}{12}) \Gamma(\frac{s}{4} - \frac{1}{12}) \Gamma(\frac{s}{4} + \frac{5}{4}) }{\Gamma(\frac{s}{4} + \frac{7}{4})}.
\end{equation}

To convert between Equations \ref{eqn:praise} and \ref{eqn:braise}, we define an equipartition factor $\eta = u_B / u_e$ as the ratio of magnetic energy density to electron energy density. Following \citet{kaiser_evolutionary_1997}, we use the lobe pressure $p = (\Gamma_c - 1)(u_e + u_B + u_T)$, where we assume the thermal energy density $u_T \sim 0$. 

We use the fluid tracer value from our numerical simulation code to define the volume filling factor of radiating particles in each Lagrangian particle packet. The tracer value is a mass tracer, defined as follows:

\begin{align}
    \begin{split}
        {\rm trc} &= \frac{n_r m_r}{n_r m_r + n_{th} m_{ th}}, \\
        \frac{n_{ th}}{n_{r}} &= \frac{m_r}{m_{th}} \left( \frac{1}{\rm trc} - 1 \right), \\
    \end{split}
\end{align}

where $n_{r}$ is the number density of radiating particles, $m_{r}$ is the mass of lobe plasma associated with each radiating particle (i.e., the sum of electron mass and positively charged particle mass), $n_{th}$ is the number density of thermal particles, and $m_{th}$ is the mass of the thermal particles. The volume filling factor then is:

\begin{equation}
\label{eqn:volumefill_BRAiSE}
    f = \frac{1}{1 + \frac{m_r}{m_{\rm th}} \left( \frac{1}{\rm trc} - 1 \right)}.
\end{equation}

We use a mass ratio of 1, corresponding to an electron-proton plasma (as $m_r = m_{th} = \mu m_p$), which simplifies this volume filling factor to a multiplication by the tracer value \citep[following][]{yates-jones_praise_2022}. The particle content of Fanaroff-Riley \citep[FR;][]{fanaroff_morphology_1974} class I and II sources is thought to be different, where FR-I sources require an extra source of internal pressure in addition to their radiating particles \citep{croston_x-ray_2005, croston_particle_2014}, indicating that FR-I-type sources may be comprised of an electron-proton plasma. In contrast, FR-II sources are likely comprised of an electron-positron plasma; however, the assumption of an electron-positron plasma results in morphological structures for FR-II-like sources that are less characteristic of structures in observed sources (see \ref{section:appendix}). We leave the exploration of different plasma compositions and implications for the particle content of AGN radio lobes for future work. 

The fluid tracer traces the mass of jet and ambient medium within each cell on the simulation grid. Within each cell, the magnetic field of the jet and ambient medium will be averaged to create a representative value. However, the emitting (jet) medium within that cell will experience a greater magnetic field strength, as the jet magnetic field is stronger than the ambient magnetic field within each cell. To account for this sub-grid effect, we use this volume filling factor to scale our magnetic field strength as $B' = B/\sqrt{f} = B/\sqrt{\rm trc}$ and magnetic field energy density as $u_B' = u_B/{\rm trc}$. Hence, we calculate our equipartition factor $\eta = u_B' / u_e$.

In addition to changing the form of the emissivity equation, we must consider the dependence of the radiative losses on the magnetic energy density. The Lorentz factor of the electron population changes over time as \citep{turner_raise_2018_II}:

\begin{equation}
    \frac{d \gamma}{dt} = \frac{a_p}{3 \Gamma_c} \frac{\gamma}{t} - \frac{4}{3} \frac{\sigma_T}{m_e c} \gamma^2 (u_B(t) + u_C),
\end{equation}

where $a_p$ is the exponent with which the pressure changes locally (i.e. $p \propto t^{a_p}$), $\sigma_T$ is the electron scattering cross-section, and the IC losses are described by $u_C = 4.00 \times 10^{-14} (1 + z)^4$ Jm$^{-3}$, where $z$ is the redshift of the radio source. \citet{turner_raise_2018_II} describe how the distribution of electron energies changes over time for arbitrary lobe expansion using a recursive relation:

\begin{equation}
    \gamma_{n} = \frac{\gamma_{n-1} t^{a_p(t_{n-1}, \; t_n)/3 \Gamma_c}_{n}}{t_{n-1}^{a_p(t_{n-1}, \; t_n)/3 \Gamma_c} - a_2(t_{n}, t_{n-1}) \gamma_{n-1}},
\end{equation}

where $t_n$ and $\gamma_n$ are the time and the corresponding Lorentz factor at the $n$th time-step. The $a_2(t_{n}, t_{n-1})$ term is given by:

\begin{equation}
    a_2 (t_{n}, t_{n-1}) = \frac{4}{3} \frac{\sigma_T}{m_e c} \left[ \frac{u_B(t_{n})}{a_3} t_{n}^{-a_B} (t_{n-1}^{a_3} - t_{n}^{a_3}) + \frac{u_C}{a_4} (t_{n-1}^{a_4} - t_{n}^{a_4}) \right],
\end{equation}

where $a_B$ is the exponent with which the magnetic energy density changes locally (i.e. $u_B \propto t^{a_B}$, $a_3 = 1 + a_p/3\Gamma_c + a_B$, and $a_4 = 1 + a_p/3\Gamma_c$. In \citet{turner_raise_2018_II} and \citet{yates-jones_praise_2022}, where the magnetic energy density is assumed to scale with the pressure, the $a_B$ exponent is replaced with the $a_P$ exponent, and the $a_3$ term is changed accordingly. 

\subsection{Stokes parameters}
\label{section:stokesparams}

Synchrotron emission is linearly polarised, and hence the Stokes parameters are given by \citep{del_zanna_simulated_2006}:

\begin{align}
\label{eqn:stokes_params}
\begin{split}
    I_\nu(x,z) &= \int^{\infty}_{-\infty} J_\nu(x, y, z) dy,\\
    Q_\nu(x,z) &= \frac{\alpha + 1}{\alpha + 5/3} \int^{\infty}_{-\infty} J_\nu(x, y, z) \cos(2 \chi_0) dy, \\
    U_\nu(x,z) &= \frac{\alpha + 1}{\alpha + 5/3} \int^{\infty}_{-\infty} J_\nu(x, y, z) \sin(2 \chi_0) dy. \\
\end{split}
\end{align}

Here, the $y$-axis is the line of sight, $\alpha$ is the injection spectral index and $\chi_0$ is the initial polarisation position angle in the plane of the sky (i.e. before Faraday rotation has been applied). We calculate $\chi_0$ following \citet{del_zanna_simulated_2006}, accounting for relativistic effects on the position angle:

\begin{align}
\label{eqn:chi0}
    \begin{split}
        \cos(2 \chi_0) &= \frac{q_x^2 - q_z^2}{q_x^2 + q_z^2}, \\
        \sin(2 \chi_0) &= \frac{-2 q_x q_z}{q_x^2 + q_z^2}, \\
    \end{split}
\end{align}

where $q_x = (1 - \beta_y)B_x + \beta_x B_y$ and $q_z = (1 - \beta_y)B_z + \beta_z B_y$, where $B_i$ are the magnetic field vector components and $\beta_i = v_i / c$ are the dimensionless velocity vector components. The fractional polarisation is then given by:

\begin{equation}
\label{eqn:fracpolar}
    \Pi = \frac{\sqrt{Q^2 + U^2}}{I}.
\end{equation}

We calculate Faraday rotation measures for our Lagrangian particles using a method similar to that described in \citet{jerrim_faraday_2024}; that is, we integrate the Faraday rotation measure ($\phi$) along the line of sight to the location of the Lagrangian particle. 

\section{Validation}
\label{section:validation}

\subsection{Simulations}
\label{section:simulations_BRAiSE}

The simulations in this paper are carried out using the \textsc{PLUTO} astrophysical fluid dynamics code \citep[version 4.3;][]{mignone_pluto_2007} using the relativistic magnetohydrodynamics (RMHD) physics and relativistic hydrodynamics (RHD) physics modules. The HLLD Riemann solver is used, as well as 2nd order Runge-Kutta timestepping with a Courant-Friedrichs-Lewy (CFL) number of $0.33$ and linear reconstruction. The $\nabla \cdot \mathbf{B} = 0$ condition is controlled by Powell's eight-wave formulation \citep{powell_approximate_1997,powell_solution-adaptive_1999}, which is used to prevent nonphysical changes in the magnetic field structure. Radiative cooling is not considered in these simulations; all radiative losses are calculated in post-processing as they do not significantly affect jet dynamics in the evolutionary stages considered in this work \citep{hardcastle_simulation_2018}.

These simulations were carried out on a three-dimensional Cartesian grid centred at $(0,0,0)$. Each dimension contains five grid patches; a central uniform grid with $100$ grid cells from $-2.5 \rightarrow +2.5$ kpc with a resolution of $0.05$ kpc/cell to ensure jet injection is sufficiently resolved; two stretched grid patches from $\pm 2.5 \rightarrow \pm 10$ kpc; and two stretched grid patches from $\pm 10 \rightarrow \pm 200$ kpc. The stretched grid patches contain $100$ and $330$ cells respectively, with typical resolutions of $0.11$ kpc/cell at $10$ kpc and $0.86$ kpc/cell at $100$ kpc. The total grid size is $960^3$ with periodic boundary conditions at each outer boundary. The unit values of length, velocity and density in our simulations are $\hat{L} = 1$ kpc, $\hat{v} = c$, and $\hat{\rho} = 1.002 \times 10^{24}$ g/cm$^3$ respectively.

Our simulations are carried out in radially averaged versions of the cluster and group environments used in the CosmoDRAGoN simulations \citep{yates-jones_cosmodragon_2023}, originally taken from \textsc{The Three Hundred} project \citep{cui_three_2018}. We refer to these environments as the radially averaged cluster (RAC) and radially averaged group (RAG) respectively. Using the method outlined in \citet{jerrim_faraday_2024}, we have generated a magnetic field for the group environment with an average magnetic field strength of $0.1 \mu$G and minimum wavenumber $k_{\rm min} = 0.015$ kpc$^{-1}$. The cluster environment uses the same magnetic field structure with an average magnetic field strength of $1 \mu$G.

To simulate FR-II like sources, we inject fast, relativistic, conical jets with total one-sided jet power $Q_j = 1 \times 10^{38}$ W, jet Lorentz factor $\Gamma = 5$, ratio of kinetic to thermal energy in the jet $\chi = 100$, and jet half-opening angle $\theta = 15\deg$. The jets are injected with 80 Lagrangian passive tracer particles every 0.01 Myr. These particles are used for the radio emissivity calculations. 

We simulate three different jets, with differing jet magnetic field strengths. We fix the total energy flux onto the grid to be constant for all four simulations and change the ratio of magnetic to kinetic momentum flux by changing the input jet magnetic field. In CGS units, the total energy flux onto the grid is given by:

\begin{align}
\label{eqn:total_E_flux_BRAiSE}
\begin{split}
    F_{j} &= \gamma^2 \rho c^2 + p \left( \frac{\Gamma}{\Gamma - 1} \gamma^2 - 1 \right) \\
    &+ \frac{1}{8 \pi} \left[ \left( \left( 1 + \left( \frac{\mathbf{v}}{c} \right)^2 \right) \mathbf{B}^2 \right) - \left( \left(\frac{\mathbf{v}}{c} \right) \cdot \mathbf{B} \right)^2 \right].
\end{split}
\end{align}

The jet setup is as in \citet{jerrim_faraday_2024}, with a toroidal jet magnetic field decreasing in strength from its initial value at the base of the injection cone. We use Equation \ref{eqn:total_E_flux_BRAiSE} to solve for the density and pressure in the jet using the magnetic field strength at the outer edge of the jet injection region.

\begin{table}
	\centering
	\caption[Simulation parameters]{Parameters of the simulations discussed in this paper. $B$ is the initial magnetic field strength in the `cap' of the injection cone. $\rho$ and $p$ are the density and pressure in the jet, respectively. $Q_{B}/Q_{k}$ is the ratio of magnetic to kinetic energy flux in the jet. The letters in the `env' column correspond to the type of environment the simulation has; `G' for group and `C' for cluster.}
	\label{tab:sim_list}
    \begin{tabular}{llllll}
    \hline
         Name & $B$ ${\rm (} \mu {\rm G)}$ & $\rho$ ${\rm (g/cm^3)}$ & $p \; {\rm (Pa)}$  & $Q_{B}/Q_{k}$ & Env \\
    \hline
         RAG-B0 & N/A & $3.94 \times 10^{-30}$ & $1.44 \times 10^{-11}$ & N/A & G \\
         RAG-B16 & $16$ & $3.94 \times 10^{-30}$ & $1.44 \times 10^{-11}$ & 0.0003 & G \\
         RAG-B327 & $327$ & $3.57 \times 10^{-30}$ & $1.30 \times 10^{-11}$ & 0.13 & G \\
         RAC-B327 & $327$ & $3.57 \times 10^{-30}$ & $1.30 \times 10^{-11}$ & 0.13 & C \\
    \hline
    \end{tabular}
\end{table}

\subsection{Dynamics}
\label{section:dynamics_BRAiSE}

The dynamics of our jets are highly dependent on the ratio of magnetic to kinetic energy flux in the jet. As described previously by \citet{mukherjee_simulating_2020}, highly magnetised jets with a toroidal field structure are very narrow and do not follow self-similar evolution. Magnetic field topology influences large-scale jet dynamics \citep[e.g.][]{koessl_numerical_1990, chen_numerical_2023}, however, the dynamics of our simulations are comparable as the magnetic field is toroidal in all cases. 

\begin{figure*}
    \centering
    \includegraphics{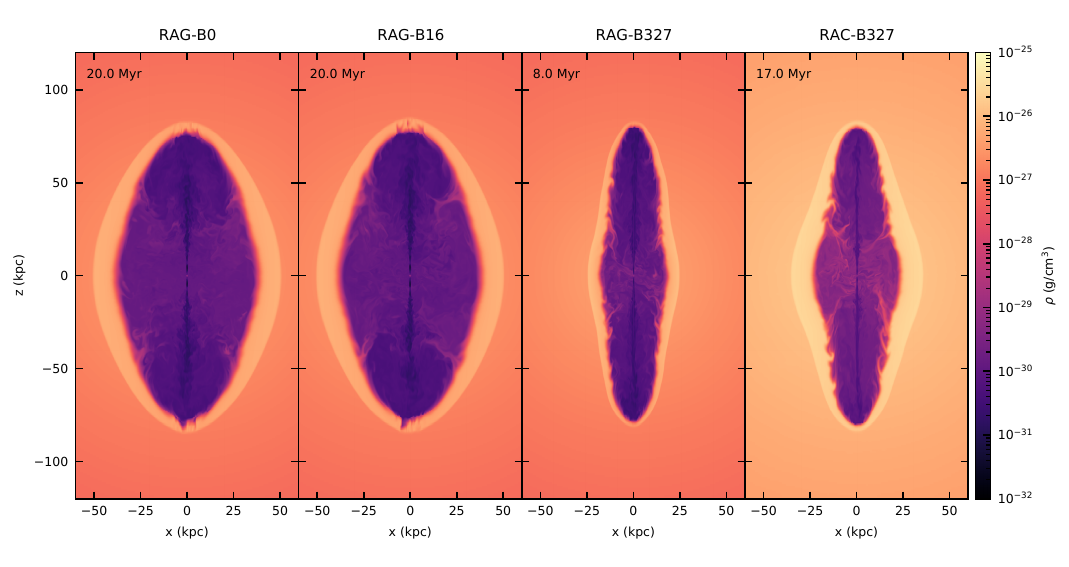}
    \caption[Midplane slices of the density at $y = 0$ kpc]{Midplane slices of the density at $y = 0$ for each simulation, with increasing jet magnetic field strength from left to right, and the cluster simulation on the far right. All four simulations are plotted at the same total radio source length of $160$ kpc. Simulations RAG-B0 and RAG-B16 are plotted at $20$ Myr. Simulation RAG-B327 is plotted at $8$ Myr. Simulation RAC-B327 is plotted at $17$ Myr.}
    \label{fig:rho-slices}
\end{figure*}

Fig. \ref{fig:rho-slices} shows this difference in the morphology of our three simulated jets, plotted at the same total lobe length. The lobes in simulations RAG-B0 and RAG-B16 are almost identical; since the jet magnetic energy is quite low ($\ll 1$ percent of the total jet power), the magnetic field does not change the dynamics of the jets in simulation RAG-B16 significantly \citep[e.g.][]{hardcastle_numerical_2014}. The lobe in simulation RAG-B327 is narrower by a factor of $2$ and reaches the total length of $160$ kpc $2.5$ times faster than simulations RAG-B0 and RAG-B16 and $2.1$ times faster than simulation RAC-B327. Since simulation RAC-B327 has a denser environment, the expansion of the lobe along the jet axis is inhibited, resulting in a narrow jet head and wide equatorial region of the lobe. 

Our simulations are comparable to the moderate-powered jets in \citet{mukherjee_simulating_2020}'s simulations D (RAG-B16) and F (RAG-B327 and RAC-B327) but shown here on larger scales as they have been evolved for longer. The difference in lobe width and jet stability is the same on these larger scales as that reported by \citet{mukherjee_simulating_2020} on galactic scales; the jet in RAG-B16 is disrupted at $0.9$ Myr by Kelvin-Helmholtz instabilities, whereas in simulations RAG-B327 and RAC-B327 the jets remain relatively stable to these instabilities due to their higher magnetisation. These simulations confirm the result of \citet{mukherjee_simulating_2020} that higher powered, faster jets, with strong magnetic fields are the most stable RMHD jets.

\begin{figure*}
    \centering
    \includegraphics{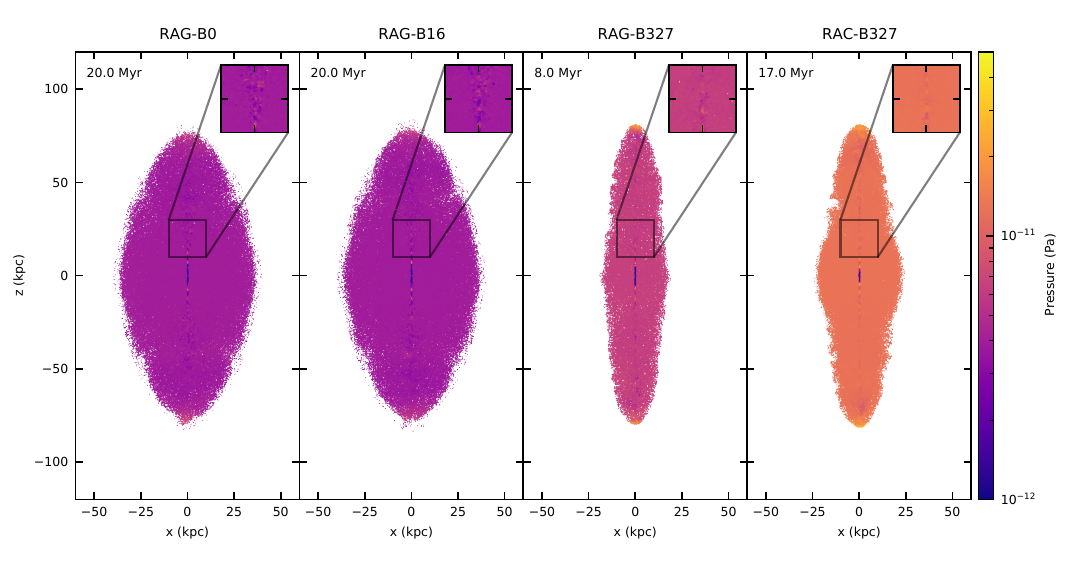}
    \caption[Scatterplot of the Lagrangian particle pressure]{Lagrangian particle pressure for each simulation, with increasing jet magnetic field strength from left to right, and the cluster simulation on the far right. The latest particles injected onto the simulation grid are plotted on top. Simulations are plotted at the same times as in Fig. \ref{fig:rho-slices}. Plot insets correspond to the region at $z \simeq 25$ kpc where the northern jets in simulations RAG-B0 and RAG-B16 decollimate.}
    \label{fig:prs-particles}
\end{figure*}

In Fig. \ref{fig:prs-particles}, we plot the pressure in the Lagrangian particles. The jets in simulations RAG-B327 and RAC-B327 terminate in a clear region of enhanced pressure. Simulations RAG-B0 and RAG-B16 are again broadly consistent with one another, with both simulations showing a small region of slightly enhanced pressures at either end of the lobes. However, the particle pressures also show the decollimation of the jet at $z \simeq \pm 25$ kpc, shown by the jet particles `fanning out' from the jet axis. This corresponds to the location of jet decollimation due to disruption, after which the jet slows down but keeps expanding outwards.

\subsubsection{Shock structures}
\label{section:shock-structures}

\begin{figure*}
    \centering
    \includegraphics{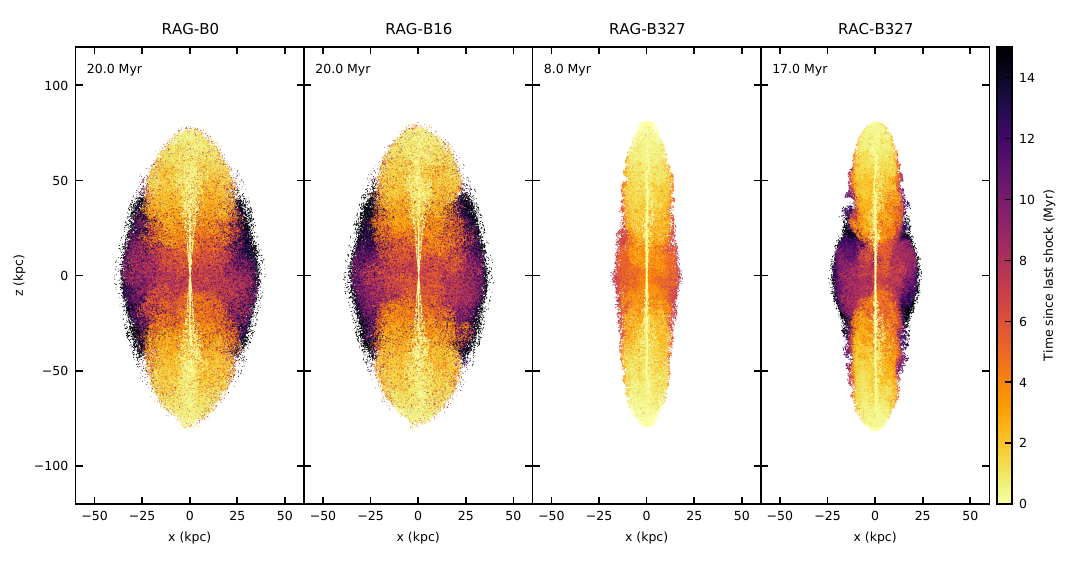}
    \caption[Scatterplot of the time since last shock for each Lagragnian particle]{Time since last shock for each Lagrangian particle plotted for each simulation, with increasing jet magnetic field strength from left to right, and the cluster simulation on the far right. The latest particles injected onto the simulation grid are plotted on top. Simulations are plotted at the same times as in Fig. \ref{fig:rho-slices}.}
    \label{fig:particle-ages}
\end{figure*}

We plot the spatial distribution of the time since each particle was either injected or last shock accelerated (i.e. particle age) in Fig. \ref{fig:particle-ages}, which shows in more detail the differences in jet stability between the three simulations. In simulations RAG-B0 and RAG-B16, we see the particle ages decrease along the jet axis, with the particles fanning out as seen in the spatial distribution of the particle pressures. The shock structures for simulations RAG-B0 and RAG-B16 are most similar to a lobed FR-I-like source. 

In comparison, simulations RAG-B327 and RAC-B327 show a clear region of freshly accelerated particles distributed over a working surface with a narrow solid angle at the jet head, which is more typical of a hotspot in an FR-II-like source. We see clear signatures of backflow with an increase in particle age from the hotspot back towards the equatorial plane. At $z = \pm 25$ kpc in both these simulations, we see some particles with an increasing age gradient flaring out from the main jet; these `flaring points' correspond to areas of Kelvin-Helmholtz instabilities along the jet. The gradient of particle ages in the backflow is concentrated over a smaller region in simulation RAC-B327 compared to simulation RAG-B327; this is due to the slower propagation of the jet, and hence the slower backflow in the cocoon.

\subsubsection{Magnetic fields}
\label{section:Bfields_BRAiSE}

\begin{figure*}
    \centering
    \includegraphics{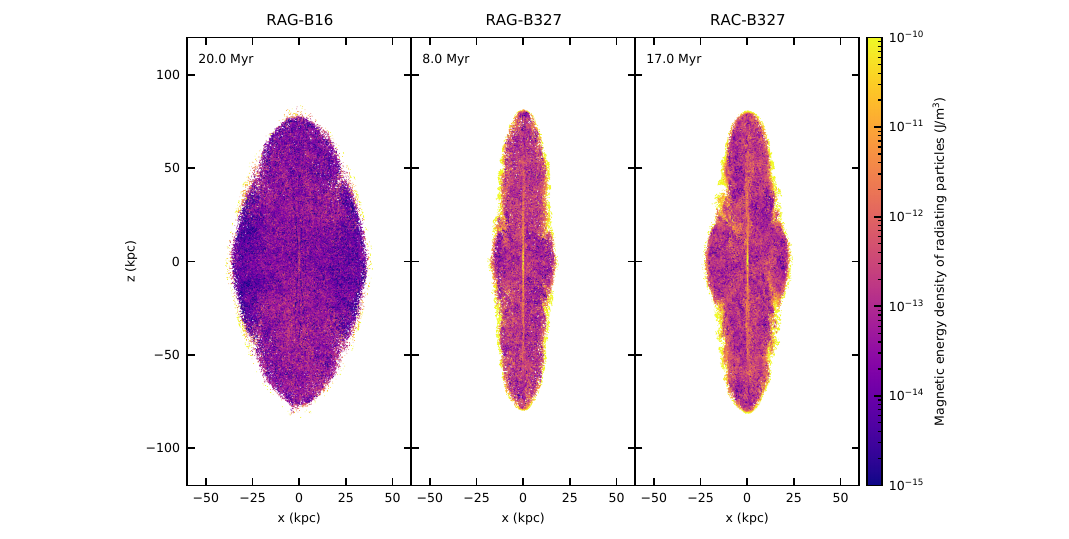}
    \caption[Scatterplot of the magnetic energy density of the radiating Lagrangian particles]{Magnetic energy density of the radiating particles for each RMHD simulation, with increasing jet magnetic field strength from left to right, and the cluster simulation on the far right. The latest particles injected onto the simulation grid are plotted on top, however, the general trends on the simulation grid are reproduced. Simulations are plotted at the same times as in Fig. \ref{fig:rho-slices}.}
    \label{fig:B-particles}
\end{figure*}

The magnetic field in the jet-inflated cocoon is highly turbulent. Fig. \ref{fig:B-particles} shows the spatial distribution of the magnetic field energy density in the particles (weighted by the fluid tracer value as discussed in Section \ref{section:synch}; this is the relevant magnetic energy for the radiating particles). The locations of strong magnetic fields will correspond to bright emission in synthetic images produced using BRAiSE. Simulation RAG-B16 shows a similar distribution of energy to the pressure; the high magnetic energy density in the jet is fanned out into the lobe, though unlike the pressure, no hotspot is seen. The magnetic energy density is generally higher towards the jet head, and lower towards the equatorial region, though in all regions it varies on small scales, reflecting the turbulent magnetic field seen in the lobe \citep[e.g. Fig. 3 of][]{jerrim_faraday_2024}.

Simulations RAG-B327 and RAC-B327 in general have higher magnetic energy densities in the jet and across the lobe for the same total source length as simulation RAG-B16 due to the higher initial jet magnetic field strength. The edges of the lobe have enhanced magnetic energy density, particularly in locations where we see Kelvin-Helmholtz instabilities in Fig. \ref{fig:rho-slices}. These regions are mixing with the shocked shell of ambient medium, which sweep up the turbulent ambient magnetic field \citep{jerrim_faraday_2024}. Like in simulation RAG-B16, the magnetic energy density is generally higher towards the jet head, and lower towards the equatorial region. Unlike the pressure (Figure \ref{fig:prs-particles}), there do not seem to be clear regions of enhanced magnetic energy density where the radio hotspots should be in both RAG-B327 and RAC-B327; this is due to kink instabilities disrupting the jet flow just before the jet head.

The differences in environment between the RAG-B327 and RAC-B327 simulations result in a slightly different distribution of high magnetic energy density regions across the cocoon. Since the cluster environment is more dense, the jet head region in simulation RAC-B327 is narrower, creating a bottleneck effect for the particles in the backflow. The particles in the backflow are swept into pockets of the cocoon formed by Kelvin-Helmholtz instabilities at the boundary between the cocoon and the shocked shell of ambient medium surrounding the cocoon. The mixing that these instabilities cause can be seen in Fig. \ref{fig:rho-slices} (e.g. at $x \simeq -10$, $z \simeq 25$), where these regions cause filaments of the higher density shocked ambient gas to move towards the centre of the cocoon. In addition to slowing down the backflow, this shocked gas also has a high magnetic field due to the compression of the turbulent ambient medium in the shocked shell \citep[e.g.][]{jerrim_faraday_2024}. The particles in these regions will therefore have a slightly higher magnetic energy density than those elsewhere in the cocoon. This non-uniform distribution of magnetic energy is a dynamical effect due to the density of the environment.

\begin{figure}
    \centering
    \includegraphics{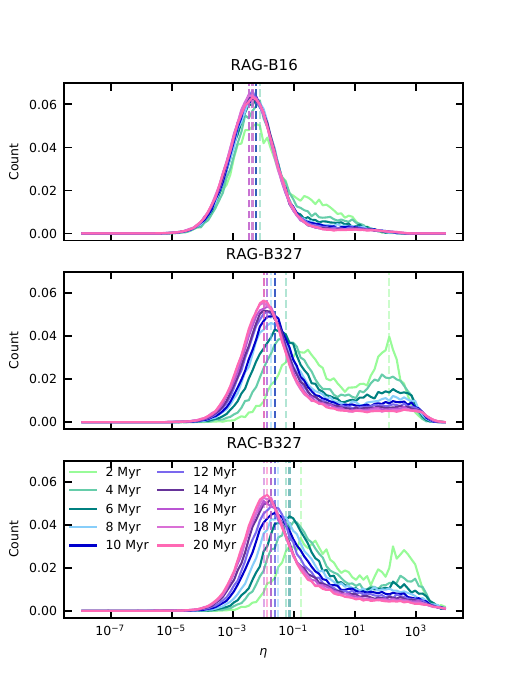}
    \caption[Equipartition factor distributions over time]{Distributions of the equipartition factor in the Lagrangian particles over time for the RMHD simulations, with increasing jet magnetic field strength from top to bottom, and the cluster simulation shown in the bottom panel. The distributions are shown for every $2$ Myr until $20$ Myr for each RMHD simulation.}
    \label{fig:eta-histograms}
\end{figure}

The non-magnetic approach to calculating radio emission requires a mapping between pressure and magnetic field strength. We determine an appropriate equipartition factor as defined by the ratio of magnetic to particle energy densities ($\eta = \frac{u_B'}{u_e}$, where $u_e$ is defined in Equation \ref{eqn:ue}, and $u_B' = u_B / {\rm trc}$) to describe the average magnetic energy density across the whole lobe for the PRAiSE calculation. Fig. \ref{fig:eta-histograms} shows the evolution of the distribution of the equipartition factor over time in the Lagrangian tracer particles for our RMHD simulations. The distributions for our simulations have peak $\eta$ values between $0.1$ and $0.001$ at most times, which is consistent with observations \citep[e.g.][]{turner_raise_2018_III}. Because the equipartition factors tend to decrease over time, we adopt the peak in the equipartition factor distribution at each time in our simulations, rather than averaging these peak values over time.

At early times in each simulation, a secondary peak at higher $\eta$ values is seen; this is a numerical artefact of how the Lagrangian particles are set with values that are interpolated from the nearest grid cells. For particles at the edges of the lobe, the tracer value is extremely low due to interpolation with grid cells outside the jet, resulting in very high magnetic energy densities, and therefore high $\eta$ values. This effect is more significant in simulation RAG-B327 due to the narrowness of the jet; for the first $2$ Myr, we take the secondary peak of the $\eta$ value distribution to describe the magnetic energy density more accurately within the lobe. These high $\eta$ value particles will not affect the overall synthetic radio emission since their emission will be downweighted by the fluid tracer value. 

\subsection{Surface brightness}
\label{section:radio_BRAiSE}

As described in Section \ref{section:braise-eqns}, we generate synthetic radio synchrotron emission including IC and adiabatic losses using either the particle pressures (PRAiSE) or the particle magnetic field strengths (BRAiSE). Following \citet{yates-jones_praise_2022}, we define the spectral index as $S_\nu \propto \nu^{-\alpha}$ and use an injection index of $\alpha_{\rm inj} = 0.6$, and minimum and maximum Lorentz factors $\gamma_{\rm min} = 500$ and $\gamma_{\rm max} = 10^5$ respectively.

\begin{figure*}
    \centering
    \includegraphics{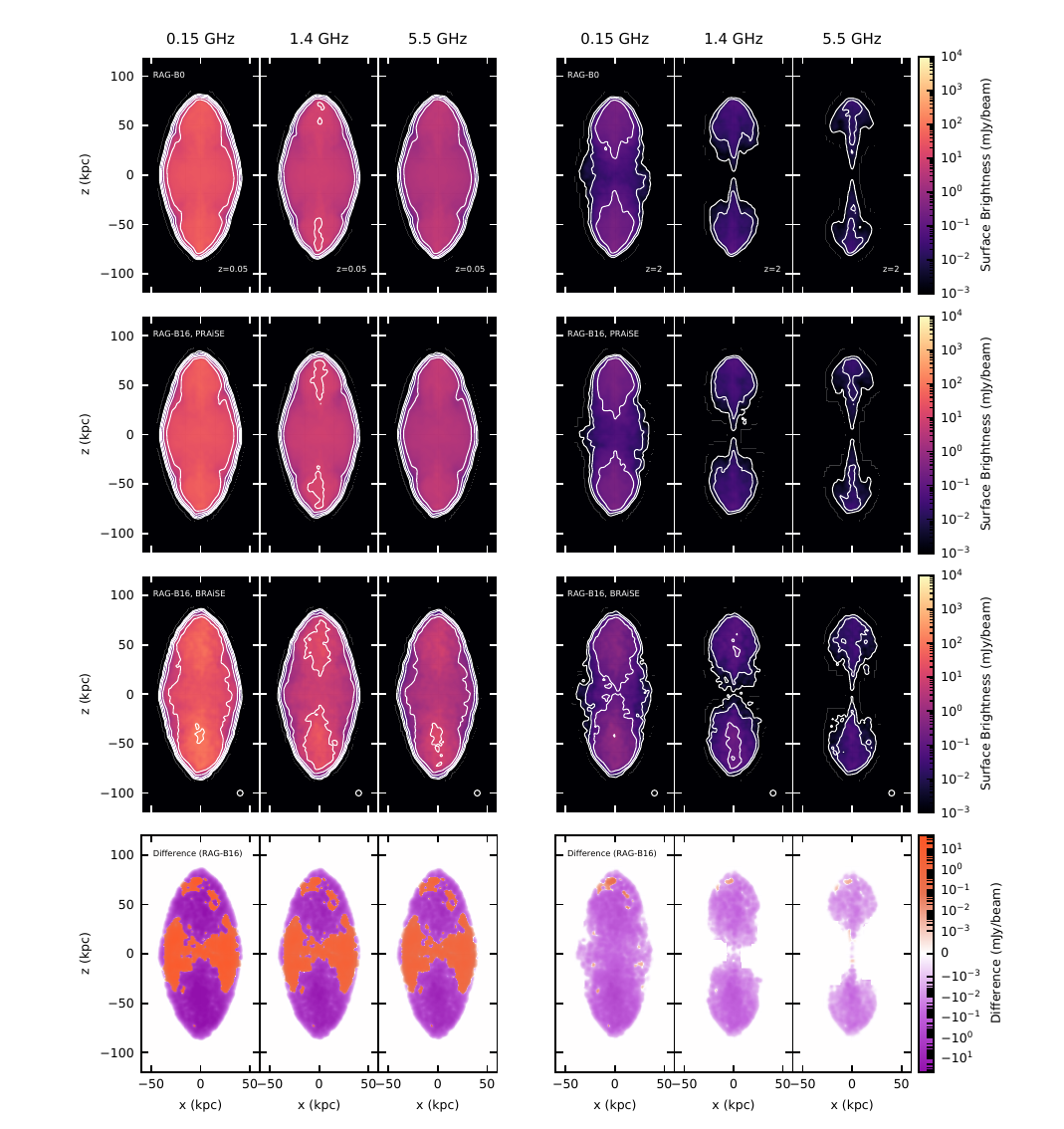}
    \caption[Surface brightness plots for simulations RAG-B0 and RAG-B16]{PRAiSE and BRAiSE surface brightnesses (Stokes I) for the RAG-B0 and RAG-B16 simulations at $20$ Myr. Contours are $0.001, 0.01, 0.1, 1, 10, 100, 1000,$ and $10000$ mJy/beam. Each surface brightness is plotted for two redshifts ($z = 0.05, 2$, as indicated in the bottom right-hand corner of the top row) and three frequencies ($0.15, 1.4,$ and $5.5$ GHz). The bottom row corresponds to the difference between the PRAiSE and BRAiSE surface brightnesses for simulation RAG-B16. The telescope observing beam is shown as a circle in the bottom right-hand corner of the panels in the BRAiSE surface brightness row.}
    \label{fig:sb-braise-B16}
\end{figure*}

\begin{figure*}
    \centering
    \includegraphics{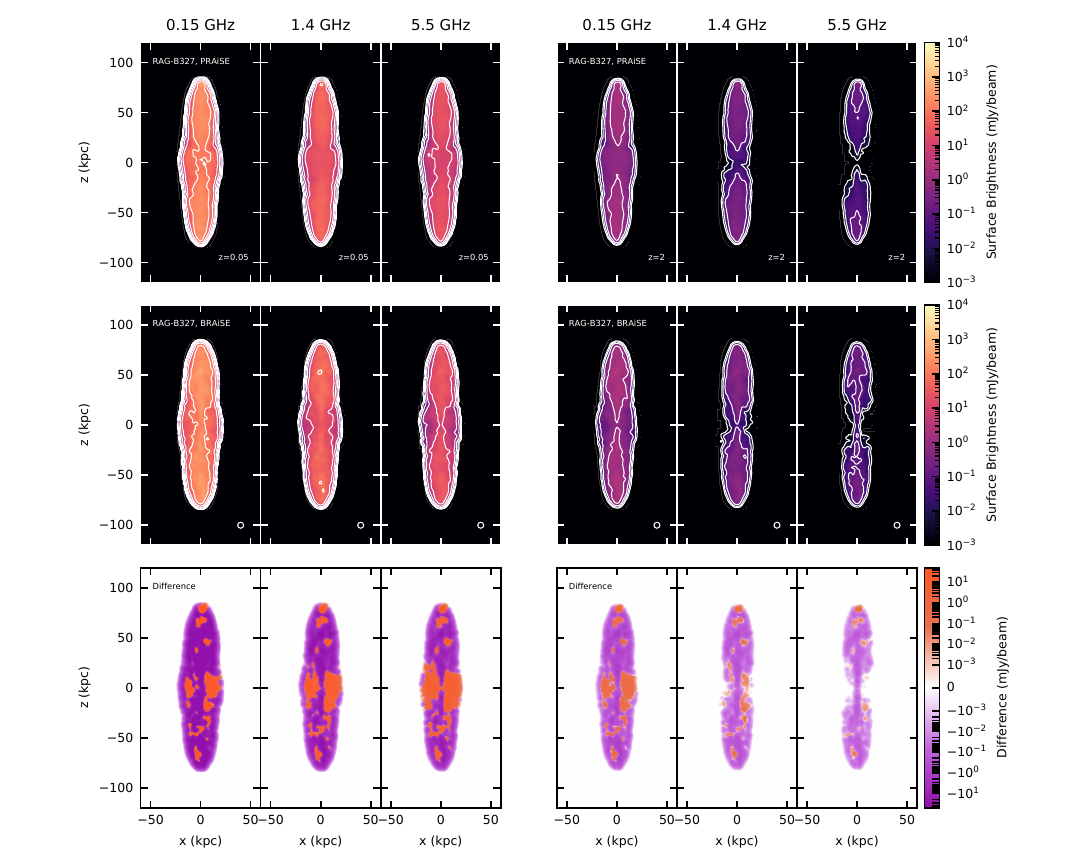}
    \caption[Surface brightness plots for simulation RAG-B327]{PRAiSE and BRAiSE surface brightnesses (Stokes' I) for the RAG-B327 simulation at $8$ Myr. Contours, redshifts, frequencies, and observing beam are as for Fig. \ref{fig:sb-braise-B16}. The bottom row corresponds to the difference between the PRAiSE and BRAiSE surface brightnesses for simulation RAG-B327.}
    \label{fig:sb-braise-B327}
\end{figure*}

\begin{figure*}
    \centering
    \includegraphics{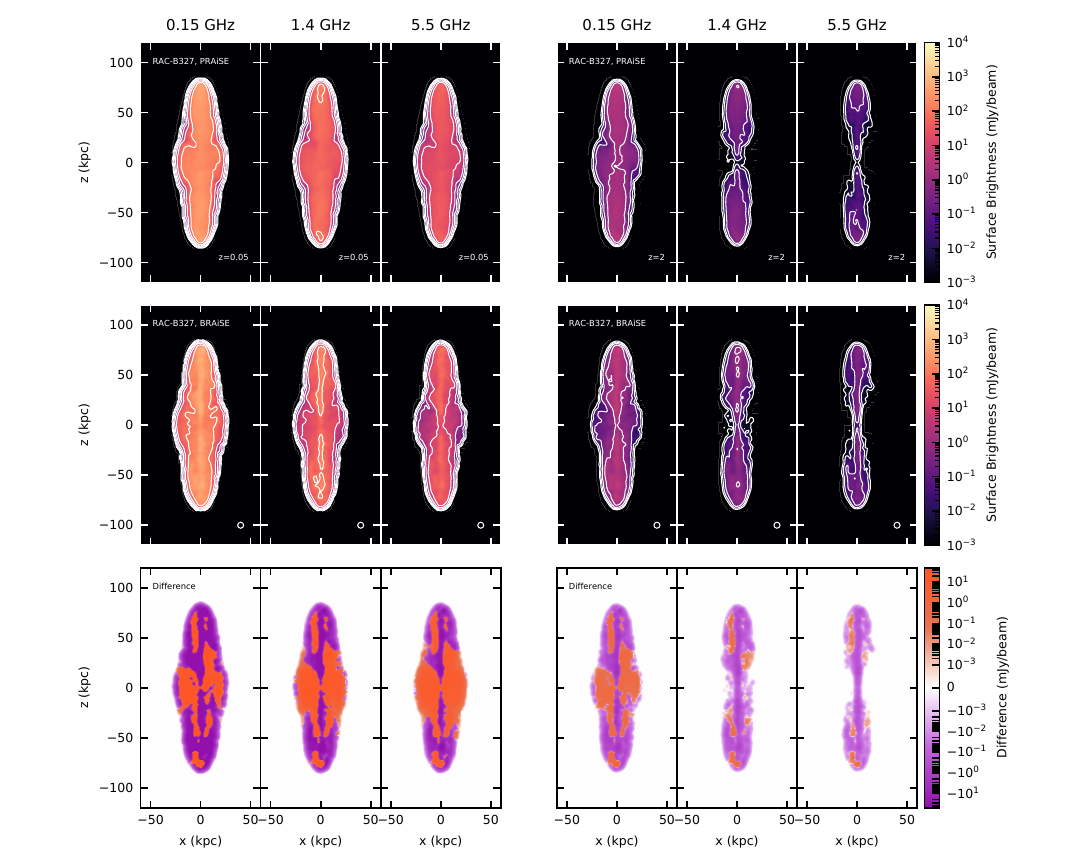}
    \caption[Surface brightness plots for simulation RAC-B327]{PRAiSE and BRAiSE surface brightnesses (Stokes' I) for the RAC-B327 simulation at $17$ Myr. Contours, redshifts, frequencies, and observing beam are as for Fig. \ref{fig:sb-braise-B16}. The bottom row corresponds to the difference between the PRAiSE and BRAiSE surface brightnesses for simulation RAC-B327.}
    \label{fig:sb-braise-RAC_B327}
\end{figure*}

We calculate the surface brightness images in the plane of the sky following the method outlined in \citep{yates-jones_praise_2022}. The images have been convolved with a 2D Gaussian beam with FWHM of $2.97''$ at a redshift of $z = 0.05$ and $0.35''$ at $z = 2$ (this corresponds to the same physical beam size of $3$ kpc). The choice of this beam FWHM is representative of high-resolution SKA pathfinder instruments \citep{morabito_subarcsecond_2022}. In Fig. \ref{fig:sb-braise-B16}, we show the surface brightness images for simulations RAG-B0 and RAG-B16 at $20$ Myr; for the latter RMHD simulation, we include both PRAiSE and BRAiSE images and the difference between them. In Fig. \ref{fig:sb-braise-B327} we plot the surface brightness images and their difference map for simulation RAG-B327 at $8$ Myr, and similarly plot these images at $17$ Myr for simulation RAC-B327 in Fig. \ref{fig:sb-braise-RAC_B327}.

For simulations RAG-B0 and RAG-B16, we see minor differences between the PRAiSE surface brightness images. The morphology is broadly consistent between the two simulations, as expected given the similarity between the dynamical properties of these simulations. This confirms that when the jet magnetic field is subdominant, the radio emission using PRAiSE is broadly consistent with the radio emission from an RHD simulation. We see clearer small-scale structure in the BRAiSE surface brightness image due to the turbulent magnetic field structure seen in Fig. \ref{fig:B-particles} \citep[as also seen in][]{tregillis_synthetic_2004, huarte-espinosa_3d_2011, hardcastle_numerical_2014}. The difference map shows that in general, BRAiSE is brighter along the jet and in the main lobe region, but PRAiSE is brighter in the equatorial region. This is due to the distribution of magnetic fields: in PRAiSE the effective lobe magnetic field is dependent on the pressure, which is relatively constant across the lobe (Fig. \ref{fig:prs-particles}) compared to the turbulent magnetic field (Fig. \ref{fig:B-particles}). As a result, the emission along the jet will be underpredicted and the emission in the equatorial regions will be overpredicted. The relative difference in emission is consistent across all frequencies at low redshift ($z = 0.05$).

For the higher jet magnetic field simulations, we again find more defined small-scale structure in the BRAiSE surface brightness image. The BRAiSE surface brightness also shows a defined cone of bright emission which is not seen in the PRAiSE image at $z \pm 25$ kpc (outlined by the highest contour at $5.5$ GHz). This is due to the presence of particles with high magnetic energies in the flaring region (as discussed in Section \ref{section:shock-structures}). This region is highly affected by Kelvin-Helmholtz instabilities that allow some of the radiating particles to escape the jet channel and flow forward into the lobe. This effect is stronger in simulation RAG-B327 than simulation RAC-B327, as the confinement of the cluster environment aids the stability of the jet in the latter simulation. This flaring region plus hotspot emission results in a bimodal surface brightness morphology that resembles hybrid sources such as J1154+513, J1206+503, and J1313+507 \citep[][]{harwood_unveiling_2020}.

For simulations RAG-B327 and RAC-B327, the jet emission using BRAiSE is brighter than in PRAiSE due to the high jet magnetic field strength which is not captured in the peak equipartition factor used in PRAiSE. Despite this enhanced jet emission, the difference map clearly shows that BRAiSE lacks defined hotspots in simulation RAG-B327. This is partially due to the kink instabilities in the jet spreading the magnetic energy (as noted in Section \ref{section:Bfields_BRAiSE}), similar to what \citet{tingay_high_2008} hypothesise is occurring in the southeast hotspot of Pictor A. This results in a larger working surface for the jet head; however, since the pressure is still enhanced in this region, we still find a hotspot-like structure when using PRAiSE. This enhanced jet emission is discussed further in Section \ref{section:FRindex}.

For simulations RAG-B327 and RAC-B327, PRAiSE has some brighter areas in the equatorial regions of the lobe that coincide with locations of lower magnetic energy density (Fig. \ref{fig:B-particles}). There are also significant differences between PRAiSE and BRAiSE in the small-scale structure for simulation RAC-B327. This is due to the distribution of magnetic energy density in the lobe (as discussed in Section \ref{section:Bfields_BRAiSE}); BRAiSE shows brighter regions of surface brightness in the areas where the backflow of particles from the hotspot has been slowed down due to the mixing of the cocoon with the shocked shell. An example of this bright region is at  $x \simeq -10$, $z \simeq 25$, corresponding to a region of significant mixing. The restricted movement of the backflow into the cocoon due to the higher density environment can be seen at high redshift ($z = 2$) for PRAiSE.

In all simulations, we find that the differences between PRAiSE and BRAiSE are consistent at a higher redshift ($z = 2$); the BRAiSE surface brightness is in general higher, meaning that more of the radio lobe is visible at higher redshifts (and frequencies). Therefore, PRAiSE produces surface brightnesses that are more susceptible to IC losses. These losses result in more `pinched' or narrower lobes for all simulations, which become more extreme at higher frequencies. In simulation RAG-B327, this effect is more subtle due to the narrowness of the lobe and brightness of the jet. At a redshift of $z = 2$, the difference maps show smaller contrasts due to the reduced magnitude of the synchrotron emission and greater importance of IC losses, which are the same for both PRAiSE and BRAiSE. 

The integrated $0.15$ GHz luminosity at a redshift of $z = 0.05$ for all four simulations are not similar to one another (Table \ref{tab:luminosity}), due the clear differences in pressure and lobe magnetic field strength at the same total lobe length. BRAiSE has higher luminosities than PRAiSE ($1.8\times$ for RAG-B16, $1.4\times$ for RAG-B327 and $1.3\times$ for RAC-B327) due to the higher magnetic energy densities used, rather than the constant equipartition factor (as discussed earlier). Simulation RAC-B327 has higher luminosities due to the increased environment density confining the jet \citep{shabala_size_2013}.

\begin{table}
	\centering
	\caption[Summary of the total 0.15 GHz luminosity]{Summary of the total 0.15 GHz luminosity for each of our simulations and method of calculating the radio emission. The right-most column shows the ratio of each luminosity to the luminosity of RAG-B0 using PRAiSE.}
	\label{tab:luminosity}
    \begin{tabular}{llll}
    \hline
         Name & Method & Luminosity (W/Hz) & Ratio to RAG-B0 \\
         \hline
         RAG-B0 & PRAiSE & $8.97 \times 10^{25}$ & 1 \\
         RAG-B16 & PRAiSE & $9.52 \times 10^{25}$ & 1.06 \\
         RAG-B16 & BRAiSE & $1.68 \times 10^{26}$ & 1.88 \\
         RAG-B327 & PRAiSE & $2.32 \times 10^{26}$ & 2.58 \\
         RAG-B327 & BRAiSE & $3.27 \times 10^{26}$ & 3.65 \\
         RAC-B327 & PRAiSE & $4.23 \times 10^{26}$ & 4.72 \\
         RAC-B327 & BRAiSE & $5.64 \times 10^{26}$ & 6.29 \\
         \hline
    \end{tabular}
\end{table}

\subsection{Radio morphology}
\label{section:FRindex}

\begin{figure}
    \centering
    \includegraphics{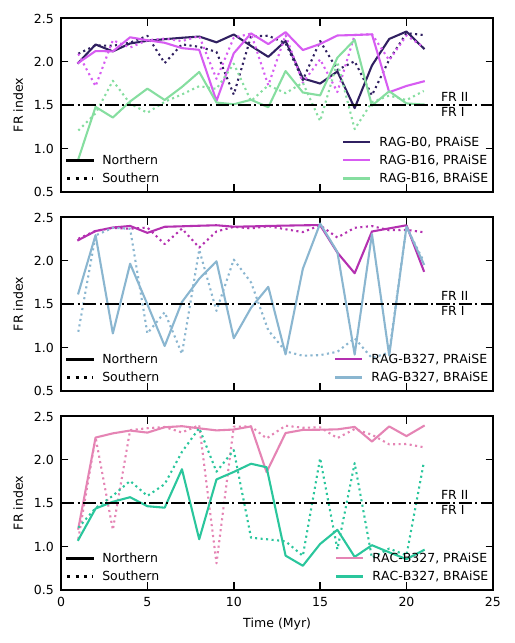}
    \caption[FR indices at $0.15$ GHz and redshift $z = 0.05$ over time]{FR indices at $0.15$ GHz and $z = 0.05$ over time for each simulation and emission code shown in Figs. \ref{fig:sb-braise-B16} (top panel), \ref{fig:sb-braise-B327} (middle panel) and \ref{fig:sb-braise-RAC_B327} (bottom panel). The classification dividing line is shown with a dot-dashed line at $FR = 1.5$. Solid lines correspond to the northern (upper) jet, and dotted lines correspond to the southern (lower) jet.}
    \label{fig:FR-index}
\end{figure}

\begin{figure}
    \centering
    \includegraphics{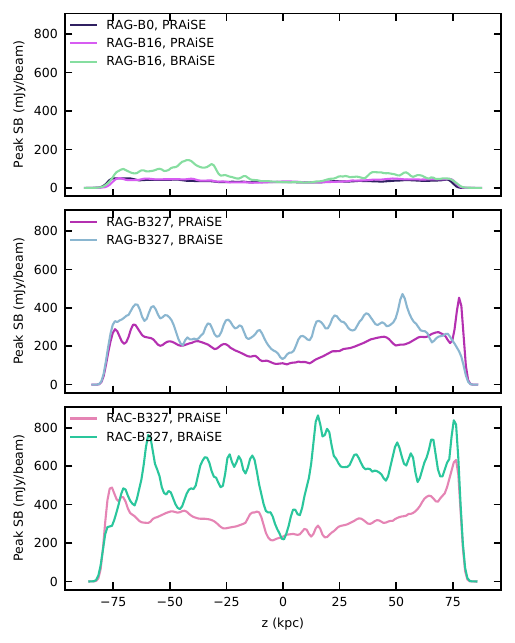}
    \caption[Peak surface brightness along the jet axis]{Peak surface brightness along the jet ($z-$)axis for each simulation and emission code at the times shown in Figs. \ref{fig:sb-braise-B16} (top panel), \ref{fig:sb-braise-B327} (middle panel) and \ref{fig:sb-braise-RAC_B327} (bottom panel).}
    \label{fig:peaksb-profile}
\end{figure}

The FR morphological classifications of our radio sources are not clearly defined. We plot the FR index over time in Fig. \ref{fig:FR-index}, following the method outlined in \citet{krause_new_2012}. The FR index is defined as ${\rm FR} = 2 x_{\rm bright}/x_{\rm length} + \frac{1}{2}$, where $x_{\rm bright}$ is the length from the origin to the brightest pixel in the surface brightness image, and $x_{\rm length}$ is the length from the origin to the furthest pixel that has a surface brightness within 2 dex of the brightest pixel. Consistent with \citet{yates-jones_praise_2022}, we calculate the FR index at $0.15$ GHz. The dividing line shown on the plot between the FR-I and FR-II classes is at $1.5$; any source with an FR index below this threshold is classed as an FR-I, and a source above the threshold is classed as an FR-II. Fig. \ref{fig:peaksb-profile} demonstrates the peak surface brightness as a function of location along the jet axis at $20$ Myr for simulation RAG-B16, $8$ Myr for simulation RAG-B327, and $17$ Myr for simulation RAC-B327.

We find for each simulation that BRAiSE produces a surface brightness image that has a lower FR index at most times in comparison to PRAiSE. The PRAiSE results for simulations RAG-B0 and RAG-B16 are broadly consistent with one another, with a time-averaged FR index across both lobes of $2.1$. However, the BRAiSE result for simulation RAG-B16 has a time-averaged FR index across both lobes of $1.6$. This is due to the brighter emission along the jet axis in BRAiSE corresponding to the regions of high magnetic energy density (Fig. \ref{fig:B-particles}). In Fig. \ref{fig:peaksb-profile}, we see that at $20$ Myr, the BRAiSE emission in both jets is brightest at approximately $52$ percent of the total jet length, unlike for the southern jet in PRAiSE, which is brightest at $90$ percent of the total jet length. This difference in location of the bright emission leads to the lower value of the FR index. As discussed in Section \ref{section:shock-structures}, this source is a lobed FR-I-like source, rather than a true FR-II. The FR indices found using the BRAiSE emission code are most consistent with this morphological classification.

The FR indices in simulation RAG-B327 are quite different for the two emission codes. The time-averaged FR index for both lobes is $2.3$ in PRAiSE, whereas it is $1.6$ in BRAiSE. In the difference maps shown in Fig. \ref{fig:sb-braise-B327}, we see clearly that the hotspots are brighter in PRAiSE. This is due to the increase in pressure at the end of the lobe; the magnetic energy is not significantly increased in this region (Fig. \ref{fig:B-particles}). At the time shown in Fig. \ref{fig:peaksb-profile} ($8$ Myr), the jets in simulation RAG-B327 are disrupted and the magnetic energy is distributed over a wider solid angle, which increases this difference in surface brightness in the hotspot region (particularly for the northern jet). It is interesting to note that there is a lack of clear classification at this time, and even as the source continues to evolve.

Simulation RAC-B327 shows very similar FR indices to simulation RAG-B327 for both PRAiSE and BRAiSE; the time-averaged FR index for both lobes is $2.2$ in PRAiSE and $1.4$ for BRAiSE. The jet emission is quite prominent in this simulation; Fig. \ref{fig:peaksb-profile} shows strong emission at $z \simeq \pm 25$ kpc, coincident with the flaring region, but also strong emission at the hotspot of the northern jet ($z \simeq 75$ kpc). On average, the flaring region is brighter than the hotspots, leading to the FR-I classification. The low ($< 1.5$) FR indices for PRAiSE at early times are due to rapid changes in direction of jet propagation, which in turn change the working surface of the jet and therefore the enhancement of the pressure in the hotspot region. This change in direction is caused by the interaction of the jet with its environment; the turbulent ambient magnetic field introduces inhomogeneities that alter the propagation of the jet.

Depending on the magnetisation of the jet and the interaction with its environment, the difference in the emission calculation method can lead to different FR classifications of the same source. The radio and dynamical classifications are inconsistent since the lobed FR-I-like sources (as identified by their shock structures) in simulations RAG-B0 and RAG-B16 produce FR-II-like radio emission (as identified by their FR indices, for both emission code methods). When using BRAiSE to include the magnetic field in the calculation of synchrotron emission, the broad morphological features of the lobes are consistent with the PRAiSE result. The emission in the jet and its associated flaring region is underpredicted for both low and high magnetisation jets in PRAiSE; this also leads to a lower overall luminosity. For high magnetisation jets, the hotspot emission is overpredicted by PRAiSE at most times, but this depends on the interaction of the jet with its environment. By including the magnetic field in the radio synchrotron emission calculation, we can also reveal details of the jet-environment interaction (e.g. simulation RAC-B327).

To ensure that these results are not an artefact of calculating the surface brightness using the Lagrangian particles, we have assessed the convergence of the calculated emissivity for different numbers of particles injected onto the simulation grid over time. We find that our particle injection rate of $80$ particles per $0.01$ Myr is more than sufficient to recover the total lobe emissivity, which is constant for an injection rate above $20$ particles per $0.01$ Myr. The flux density of the fast jet slowly decreases as more particles are added, however, this contributes a small fraction of the total emission. We note that local brightness inhomogeneities in the surface brightness images are smoothed out somewhat as the number of particles increase, and we caution against over-interpreting detailed structure in the surface brightness.\\\\\\\\

\section{Polarisation}
\label{section:polarisation_BRAiSE}

\subsection{Fractional polarisation without Faraday rotation}
\label{section:polar-noRM}

\begin{figure*}
    \centering
    \includegraphics{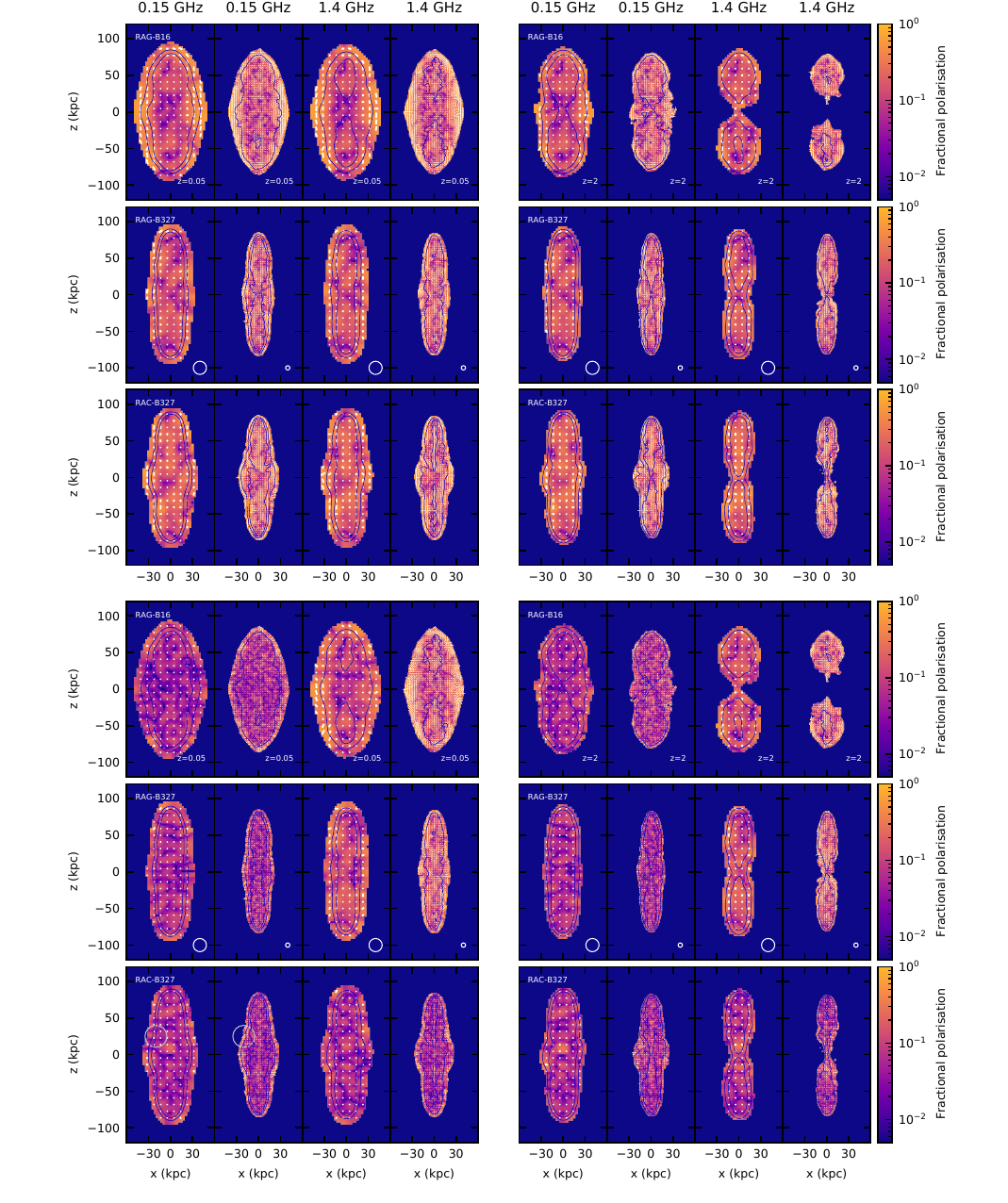}
    \caption[Fractional polarisation maps with and without Faraday rotation]{Fractional polarisation maps for the RMHD simulations using BRAiSE. The length of the vectors corresponds to the magnitude of the fractional polarisation at each location and the direction of the vectors indicates the `magnetic field direction'. Each fractional polarisation map is plotted for two redshifts ($z = 0.05, 2$, for the left- and right-hand groups respectively) and two resolutions (beam FWHM $= 3, 9$ kpc, shown as circles in the bottom right-hand corner in the middle row). The upper and lower groups correspond to the raw polarisation images and the depolarised images respectively. For the depolarised images of RAC-B327 at $z = 0.05$ and $0.15$ GHz, the circles centrered at $x = -20$, $z = 25$ correspond to the location discussed in Section \ref{section:RM}. Contours at $z = 0.05$ are at $10$ and $100$ mJy/beam, and contours at $z = 2$ are at $0.1$ and $1$ mJy/beam.}
    \label{fig:frac-polar}
\end{figure*}

The differences in jet dynamics due to jet magnetisation impact the polarisation properties of our simulated sources. In Fig. \ref{fig:frac-polar} we plot the fractional polarisation overlaid with contours corresponding to the Stokes I emission (from Figs. \ref{fig:sb-braise-B16}, \ref{fig:sb-braise-B327}, \ref{fig:sb-braise-RAC_B327}) for each of our RMHD simulations. The length of the vectors corresponds to the fractional polarisation and the direction indicates the inferred `magnetic field direction', $\chi_B = \frac{1}{2} \arctan{U/Q} + \frac{\pi}{2}$, following \citet{hardcastle_numerical_2014}. Two frequencies are plotted ($0.15$ GHz and $1.4$ GHz), at low resolution (beam FWHM of $9$ kpc, corresponding to $8.91''$ at $z = 0.05$ and $1.05''$ at $z = 2$) and high resolution (beam FWHM of $3$ kpc, corresponding to $2.97''$ at $z = 0.05$ and $0.35''$ at $z = 2$) in the left- and right-hand columns for each frequency respectively. The low resolution beam FWHM is consistent with beam FWHMs in current and upcoming polarisation surveys, which typically range between $1 - 15''$ \citep[see Table 3 of][]{jerrim_faraday_2024}. For each frequency and resolution, we plot low ($z = 0.05$) and high ($z = 2$) redshift images in the first four and latter four columns respectively. We have held the physical pixel and beam sizes constant across redshifts to compare the effect of IC losses on the fractional polarisation. In the upper three rows, we plot the raw fractional polarisation images, and in the lower three rows, we plot the depolarised by Faraday rotation fractional polarisation images (discussed in Section \ref{section:RM} below).

For each simulation, across all frequencies, redshifts, and resolutions, we find a patchy fractional polarisation structure. The fractional polarisation is higher towards the edges (on average roughly $50$ percent), a feature that is consistent with observations of polarised radio lobes \citep{hogbom_study_1979, bridle_deep_1994}. This structure is smoothed out across the source in the lower resolution images, losing much of the detail given by the complex magnetic field present in the lobes. This also reduces the magnitude of the fractional polarisation through beam depolarisation.

The amount of beam depolarisation depends on the size of the patchy fractional polarisation structure and whether that is smaller than the beam size. In simulation RAG-B16, this depolarisation occurs primarily along the jet axis, where the patchy structure is clearly seen at high resolution. For simulation RAG-B327, this depolarisation effect is strongest near the origin, while another clear patch of beam depolarisation can be seen at about $40$ kpc along the upper jet axis in Fig. \ref{fig:frac-polar}. The characteristics of the beam depolarisation are similar in simulation RAC-B327, but a clear example of this effect occurs at $x \simeq -10$, $z \simeq 25$, in a bright and highly polarised region (at high resolution) where the particles are being slowed down in the cocoon (as discussed in Section \ref{section:radio_BRAiSE}).

In each simulation, we see that many of the vectors perpendicular to the polarisation angle (i.e. the `magnetic field direction') are aligned primarily parallel to the jet axis, consistent with the structures seen in observations of radio lobes \citep{laing_model_1980, gilbert_high-resolution_2004}. Simulations RAG-B327 and RAC-B327 do not show this feature as clearly due to the narrowness of their lobes as compared to the lobes in simulation RAG-B16. The magnetic field direction in simulation RAG-B16 is in general tangential to the lobe edge, a feature that has previously been found in simulations \citep{huarte-espinosa_3d_2011}. The narrow jets in simulations RAG-B327 and RAC-B327 do not show this feature at low resolution due to beam depolarisation.

At high resolution, we resolve more changes in the magnetic field direction on small scales. In each simulation, there is higher fractional polarisation at the edges of the lobes due to less smoothing of the patchy fractional polarisation structure by beam depolarisation. The magnetic field direction follows the jet flow in general, another feature commonly seen in radio lobe observations \citep[e.g.][]{eilek_magnetic_2002, mullin_high-resolution_2006, sebokolodi_wideband_2020}. The regions of high fractional polarisation, which are primarily at the lobe edges, do not correspond to the regions of bright Stokes I emission. 

For simulation RAG-B16, we see that the fractional polarisation increases in the jet head region, as seen in observations of FR-II sources \citep{gilbert_high-resolution_2004, osullivan_untangling_2018, mahatma_low-frequency_2021}. For simulation RAG-B327, we see an increase in fractional polarisation towards the northern jet head, however, as discussed in Section \ref{section:radio_BRAiSE}, the jet head region in this simulation is not clearly defined due to the destabilisation of the jet. We do not see an increase in fractional polarisation in the hotspot region for the southern jet in this simulation. Similarly, simulation RAC-B327 does not appear to show an increase in the fractional polarisation in the jet head region.

In the higher redshift images in the four columns on the right-hand side of Fig. \ref{fig:frac-polar}, we see less emission due to IC losses (as discussed in Section \ref{section:radio_BRAiSE}). In simulation RAG-B16 in particular, we see less emission from the high fractional polarisation lobe edges that were present at low redshift. The visible emission near the jet origin has a higher fractional polarisation than at low redshift for a low resolution, with a change in the magnetic field direction to be roughly $45\degree$ from the jet axis, particularly in the northern jet. These changes with redshift are due to IC losses reducing the synchrotron emission in the equatorial region of the lobe and increasing the contribution from particles in the jet with high $u_B$. This increases the overall fractional polarisation as the jet magnetic field is highly ordered compared to the lobe magnetic field.

In the higher jet magnetic field simulations there are also higher fractional polarisation values near the origin at higher redshift, however, this difference in fractional polarisation is small compared to the difference seen for simulation RAG-B16. This is due to the strong synchrotron emission from the jet that is not as impacted by IC losses compared to the jet in simulation RAG-B16. The higher frequency of $1.4$ GHz shows greater losses for simulations RAG-B327 and RAC-B327, which again changes the magnetic field direction in the regions closest to the jet origin. For simulation RAC-B327, the magnetic field direction remains mostly unchanged, as seen in the region near $0 \lesssim x \lesssim 15, z \simeq -20$ kpc.

\begin{figure}
    \centering
    \includegraphics{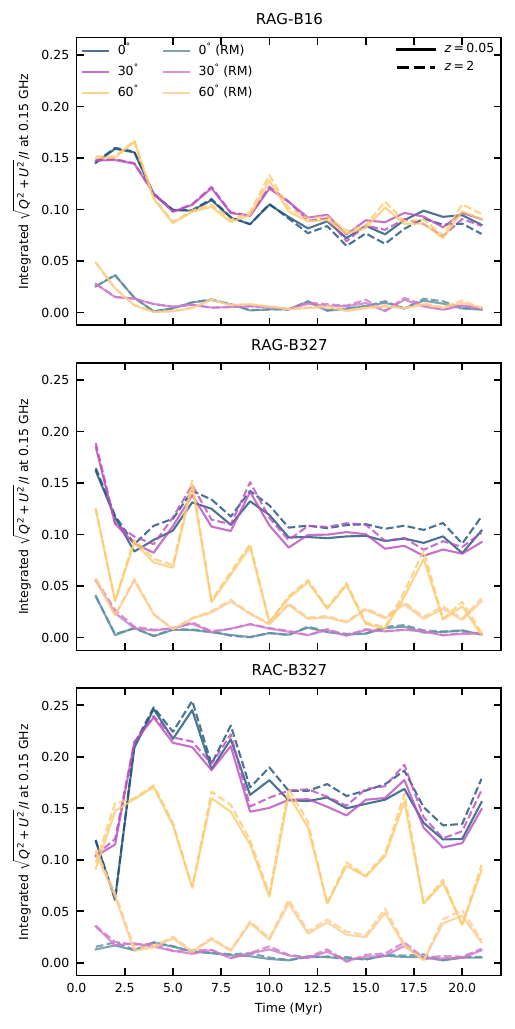}
    \caption[Integrated fractional polarisation over time]{Evolution of the integrated fractional polarisation over time for simulations RAG-B16 (top panel), RAG-B327 (middle panel), and RAC-B327 (bottom panel). Each simulation is plotted for three viewing angles ($0\degree, 30\degree, 60\degree$) and two redshifts ($z = 0.05, 2$). In the more saturated colours on the plot, we ignore the effects of Faraday rotation; Faraday rotation is applied for the desaturated colours.}
    \label{fig:time-intfracpolar}
\end{figure}

In Fig. \ref{fig:time-intfracpolar}, we plot the integrated fractional polarisation, $F_{\rm tot} = \sqrt{Q_{\rm tot}^2 + U_{\rm tot}^2}/I_{\rm tot}$; i.e. the fractional polarisation that would be measured if the source was unresolved. We include three different viewing angles to the source: $0\degree, 30\degree,$ and $60\degree$ to the plane of the sky. In this section, we do not consider the effects of Faraday rotation (these effects are discussed in Section \ref{section:RM}). 

$F_{\rm tot}$ for our simulated sources are in general below $30$ percent at all simulation times. For simulations in the group environment, $F_{\rm tot}$ initially decreases quickly (from up to $20$ percent in simulation RAG-B327) for all viewing angles. This initial decline corresponds to the lobe formation phase, where the source is dominated by jet emission, which is highly polarised due to its coherent toroidal magnetic field and is less impacted by beam depolarisation. The redshift of the source does not significantly impact $F_{\rm tot}$; for the higher jet magnetic field strength simulations, there appears to be a separation where the higher redshift sources have greater $F_{\rm tot}$, however, this difference is negligible. 

The behaviour after this initial decrease is different between each of the simulations. $F_{\rm tot}$ in simulation RAG-B16 plateaus smoothly and remains around $9 \pm 1$ percent after $5$ Myr for all viewing angles and redshifts. Simulations RAG-B327 and RAC-B327 both have an increase between $2$ and $5$ Myr before plateauing, however, this increase is greater for the cluster simulation. After $5$ Myr, at viewing angles of $0\degree$ and $30\degree$, simulation RAG-B327 has values of $F_{\rm tot}$ around $11 \pm 2$ percent, compared to $17 \pm 3$ percent for simulation RAC-B327, indicating that the more confined jet in the cluster environment has a higher intrinsic polarisation once the large-scale radio lobe is fully formed. However, in observations, this will be offset by the greater depolarising effect of the cluster environment (as discussed in Section \ref{section:RM}).

For the higher jet magnetic field strength simulations, the behaviour at an oblique viewing angle of $60\degree$ is similar to each other, in that $F_{\rm tot}$ is trending downwards with great scatter (with values ranging between $5 \pm 4$ percent for simulation RAG-B327 and $11 \pm 4$ percent for simulation RAC-B327). These sources appear quite compact at this oblique viewing angle due to the narrowness of the jet cocoon, which increases the visible amount of the lobe edge \citep[which has high fractional polarisation;][]{laing_model_1980} in comparison to the total source size. This high fractional polarisation gets averaged out in the integration across $I, Q,$ and $U$ when calculating $F_{\rm tot}$.

Despite each simulation showing a clear initial decrease in $F_{\rm tot}$, the initial value is generally lower than for past polarisation estimates from simulations \citep{hardcastle_numerical_2014, english_numerical_2016}. We also find much smaller changes in the integrated fractional polarisation over time; both \citet{hardcastle_numerical_2014} and \citet{english_numerical_2016} find a decrease from the initial maximum fractional polarisation of approximately $70$ percent to $0$ percent for lobes that evolve to $300$ kpc scales. The higher $F_{\rm tot}$ in simulation RAC-B327 indicates that a denser environment can lead to higher intrinsic polarisation of the source; the simulations discussed in \citet{hardcastle_numerical_2014} and \citet{english_numerical_2016} have a greater central density of $3 \times 10^{-26}$ g/cm$^3$ compared to $1.5 \times 10^{-26}$ g/cm$^3$ and $4.2 \times 10^{-27}$ g/cm$^3$ in our cluster and group environments respectively. This difference in environment is likely the main cause of the difference in the integrated fractional polarisation over time, however, the difference in how the polarised emission is calculated will also contribute (see Section \ref{section:stokes-comparison}).

\begin{figure}
    \centering
    \includegraphics{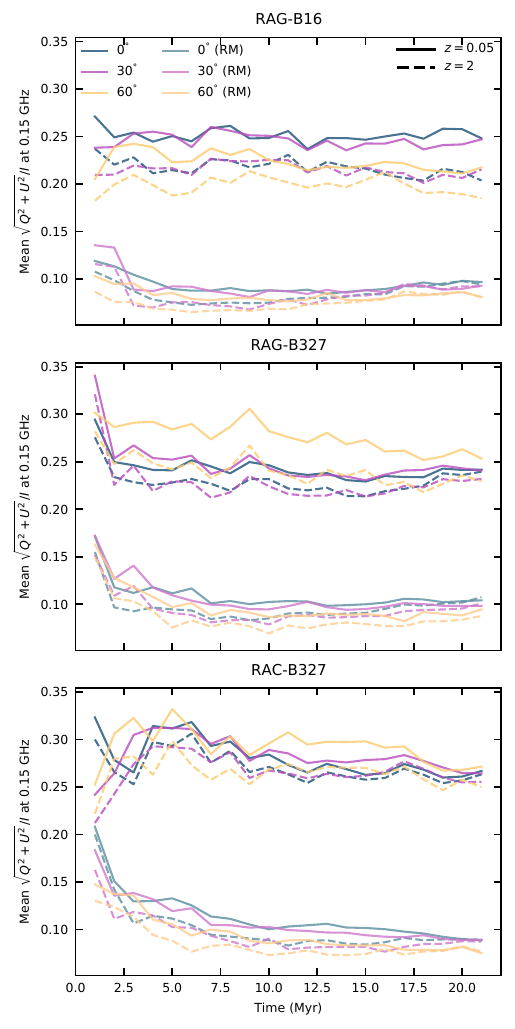}
    \caption[Mean fractional polarisation over time]{Evolution of the mean fractional polarisation over time for simulations RAG-B16 (top panel), RAG-B327 (middle panel), and RAC-B327 (bottom panel). Each simulation is plotted for three viewing angles ($0\degree, 30\degree, 60\degree$) and two redshifts ($z = 0.05, 2$). In the more saturated colours on the plot, we ignore the effects of Faraday rotation; Faraday rotation is applied for the desaturated colours.}
    \label{fig:time-meanfracpolar}
\end{figure}

In contrast, we find in Fig. \ref{fig:time-meanfracpolar} that the mean fractional polarisation, i.e. the mean value of $F = \sqrt{Q^2 + U^2}/I$ (as plotted in Fig. \ref{fig:frac-polar}), shows less evolution over time. We find that redshift and viewing angle impact the mean fractional polarisation for all our simulations. Like for the integrated fractional polarisation plot, we leave the discussion of the effects of Faraday rotation for Section \ref{section:RM}. For simulation RAG-B16, we see that the mean fractional polarisation is higher at lower redshifts for all viewing angles. When viewed at an angle of $60\degree$, the mean fractional polarisation is generally lower than when the source is viewed at a less oblique angle.

In simulations RAG-B327 and RAC-B327, we have an initial decrease of the mean fractional polarisation in the first $2$ Myr, followed by a plateau; this corresponds to the lobe formation phase. In contrast to simulation RAG-B16, we find that for the $60\degree$ viewing angle the mean fractional polarisation is in general the same or higher, e.g. the low-redshift case in simulation RAG-B327. This increase with viewing angle compared to simulation RAG-B16 is due to the morphology of the highly magnetised sources; due to their narrowness, when they are viewed at an oblique angle, the edges of the lobe (which in general have the highest fractional polarisation) contribute more to increasing the mean fractional polarisation. For simulation RAG-B327 we also see the clear difference in redshift seen for simulation RAG-B16 at a viewing angle of $60\degree$, however, it is not present for less oblique viewing angles or in simulation RAC-B327. 

\citet{hardcastle_numerical_2014} and \citet{english_numerical_2016} also find that the mean fractional polarisation evolves more slowly over time compared to $F_{\rm tot}$; however, these authors find much greater decreases over time. We also note that the mean fractional polarisations found by these authors do not drop below $30$ percent at any point, indicating that the BRAiSE method in general produces sources with lower fractional polarisation compared to these previous estimates. As discussed earlier, this is likely due to a combination of differences in the environment and the polarisation calculation for these simulated sources. In contrast, the mean fractional polarisation evolution found by \citet{meenakshi_polarization_2023} decreases from approximately $68$ percent to $36$ percent after $0.6$ Myr, which is consistent with the initial sharp decrease we find in simulation RAG-B327. Since the simulations performed by \citet{meenakshi_polarization_2023} are only run on short timescales in comparison to this work, and follow much smaller lobe sizes, we would expect that their mean fractional polarisation would continue to decrease as we have shown for simulations RAG-B327 and RAC-B327.

\subsection{Depolarisation due to Faraday rotation}
\label{section:RM}

The polarised emission from AGN jet sources is depolarised by Faraday rotation, which depends on the environment the source resides in \citep[e.g.][]{jerrim_faraday_2024}. We calculate the Faraday rotation measure (RM) by integrating Equation \ref{eqn:RM_BRAiSE} along a rotated grid for each Lagrangian tracer particle on the simulation grid, using the grid fluid variables, following the method outlined in \citet{jerrim_faraday_2024}. We assign the final RM value to each particle, which is then used to rotate the polarisation angle using Equation \ref{eqn:chi}, where the initial polarisation angle is found following Equation \ref{eqn:chi0}. This method accounts for both external and internal Faraday rotation, which will result in Faraday depolarisation.

\begin{figure}
    \centering
    \includegraphics{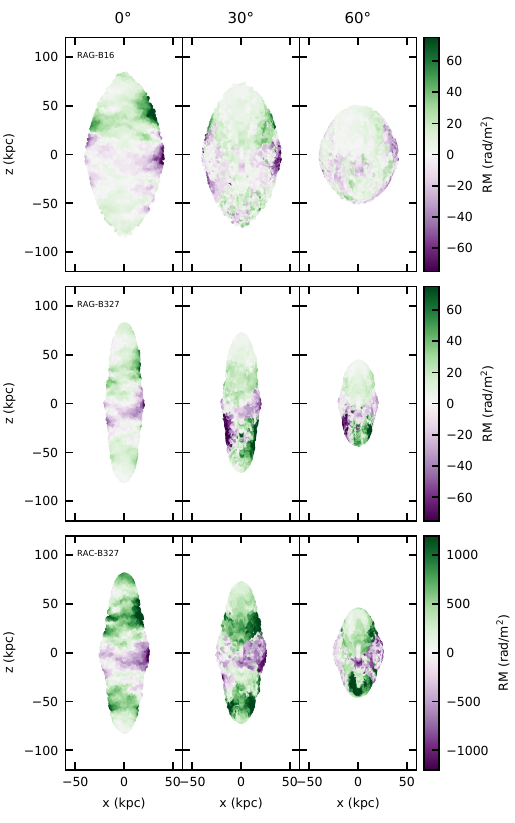}
    \caption[Scatterplot of the RM values for the Lagrangian particles]{Particle RM maps for the RMHD simulations. Each simulation is plotted with three viewing angles ($0\degree, 30\degree, 60\degree$). Top row: simulation RAG-B16. Middle row: simulation RAG-B327. Bottom row: simulation RAC-B327.}
    \label{fig:RM-maps}
\end{figure}

The Faraday rotation measures for each particle are shown in Fig. \ref{fig:RM-maps}. Simulation RAC-B327 has much greater RM values than the group simulations due to the higher density and magnetic field strength in the cluster environment. Particles towards the edges of the lobes generally have higher RM values; this is due to the alignment of the magnetic field that is pushed out by the jet cocoon with the line of sight \citep{huarte-espinosa_interaction_2011}. Otherwise, since the magnetic field structure in the ambient environment is the same, there are no major differences in the spatial structures shown in these RM maps. We note that as the viewing angle to the source is increased, the difference in RM magnitude between the upper (closer) jet and lower (further away) jet changes significantly.

We plot the fractional polarisation with Faraday rotation applied in the lower half of Fig. \ref{fig:frac-polar}. There is a clear reduction in the fractional polarisation across each map for each simulation due to Faraday depolarisation at $0.15$ GHz. The patchy structures have smaller scales, which is particularly noticeable for the higher image resolution. In simulations RAG-B16 and RAC-B327 at high redshift, we still see the region of increased fractional polarisation near the origin. The vectors perpendicular to the polarisation angle (i.e. what the inferred `magnetic field direction' is for no Faraday rotation in the high-frequency limit) are no longer tangential to the lobe edges in each simulation; in general, there is no consistency with the magnetic field directions seen in the upper half of Fig. \ref{fig:frac-polar}. The polarised emission in the group simulations (RAG-B16 and RAG-B327) is not as strongly depolarised as in the cluster simulation (RAC-B327).

At $1.4$ GHz the fractional polarisation is slightly reduced for the group simulations (as compared to Fig. \ref{fig:frac-polar}), but this effect is more subtle than for the lower frequency. Simulation RAC-B327 shows significant depolarisation at $1.4$ GHz. Since this simulation has a cluster environment with both a higher density and a higher average cluster magnetic field strength, the RMs have a wider range of values and hence the depolarisation effect is stronger at higher frequencies, as predicted in \citet{jerrim_faraday_2024}. Simulation RAC-B327 has higher fractional polarisation areas at the edges of the lobe close to where the lobe is mixing with the shocked ambient medium (e.g. at $x \simeq -20$, $z \simeq 25$), since the magnetic field at the edge of the lobe will be preferentially aligned with the line of sight due to the geometry of the jet cocoon, which is amplified more for wider cocoons \citep{huarte-espinosa_interaction_2011}. This effect is lessened in simulation RAG-B327 due to its narrow cocoon.

Both the integrated and the mean fractional polarisation over time is reduced when the RM is applied (see the desaturated lines in Figs. \ref{fig:time-intfracpolar}, \ref{fig:time-meanfracpolar}). The integrated fractional polarisation in simulation RAG-B16 falls below $5$ percent within the first few Myr, indicating that this source is almost entirely unpolarised at $0.15$ GHz. In simulations RAG-B327 and RAC-B327 for a viewing angle of $60\degree$, the integrated fractional polarisation over time is higher than that for smaller viewing angles. This behaviour is the opposite to that seen without Faraday depolarisation. At viewing angles of $0\degree$ and $30\degree$, the source is mostly unpolarised. This dramatic change in the polarisation is because the northern jet in both simulations is not strongly depolarised by Faraday rotation due to its comparatively lower RMs than the southern jet (see Fig. \ref{fig:RM-maps}), meaning the overall fractional polarisation is higher. In simulation RAC-B327, the greater RMs result in greater depolarisation of the jet emission than in simulation RAG-B327, however, the lobe edges in the cluster simulation have greater fractional polarisation values due to the alignment of the magnetic field with the cocoon. 

The mean fractional polarisation in all our simulations remains consistent after the initial decrease. After the lobe formation phase is complete ($> 5$ Myr), the mean fractional polarisation over all viewing angles and redshifts in simulations RAG-B16, RAG-B327, and RAC-B327 drops from $22 \pm 2$ percent, $24 \pm 2$ percent and $27 \pm 2$ percent without Faraday depolarisation to $8.3 \pm 0.8$ percent, $9.2 \pm 0.9$ percent and $9 \pm 1$ percent when Faraday depolarisation is applied. The decrease in mean fractional polarisation is greatest for simulation RAC-B327 due to its higher density cluster environment and therefore higher RM values \citep[e.g.][]{jerrim_faraday_2024}. 

\citet{hardcastle_numerical_2014} study the Laing-Garrington effect in their simulations using a depolarisation ratio (DPR), measured as a ratio of the depolarisation of the jet and counterjet. This depolarisation for each jet is the ratio of polarised flux density (with Faraday rotation applied) for a low and high frequency respectively; \citet{hardcastle_numerical_2014} choose frequencies of $1.4$ and $5.0$ GHz. The DPR should be greater than 1 if the jet (pointed towards the observer) is less depolarised than its counterjet (pointed away from the observer), as expected from the Laing-Garrington effect \citep{laing_sidedness_1988, garrington_systematic_1988}. This is due to the fact that the jet that is pointed towards the observer will have lower Faraday rotation measure values as there will be a shorter path length through the shocked shell of ambient medium that surrounds the jet cocoon \citep[e.g.][]{jerrim_faraday_2024}. 

\begin{figure}
    \centering
    \includegraphics{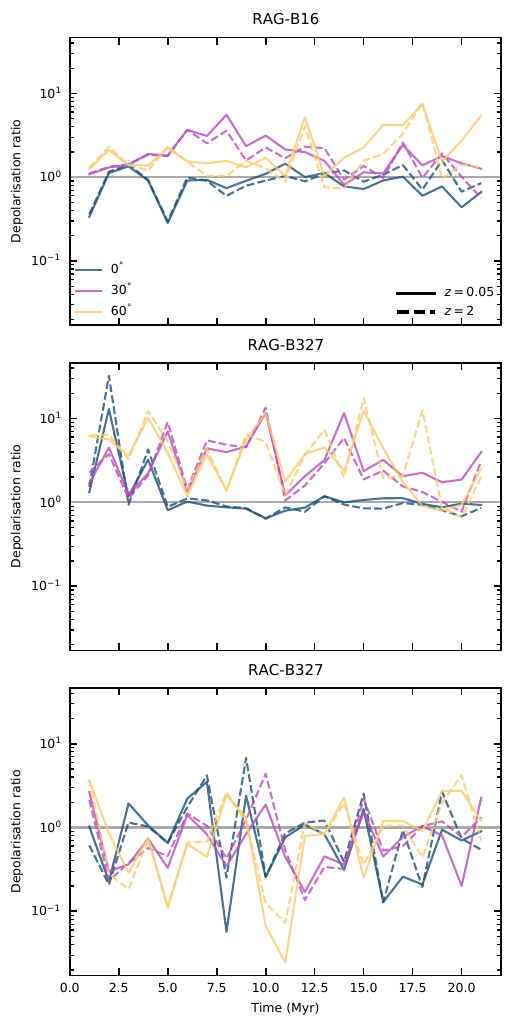}
    \caption[Depolarisation ratio over time]{Evolution of the depolarisation ratio over time for simulations RAG-B16 (top panel), RAG-B327 (middle panel), and RAC-B327 (bottom panel). Each simulation is plotted for three viewing angles ($0\degree, 30\degree, 60\degree$) and two redshifts ($z = 0.05, 2$). The horizontal grey line indicates a DPR of 1.}
    \label{fig:time-depolarratio}
\end{figure}

We plot the DPR for $1.4$ and $5.5$ GHz in Fig. \ref{fig:time-depolarratio}. For viewing angles $> 0 \degree$ in our group simulations, the DPR is in general $> 1$, consistent with our expectations from the differences in RM magnitudes (Fig. \ref{fig:RM-maps}) and the Laing-Garrington effect. However, the DPR is not always $> 1$ due to the differences in the turbulent magnetic field structure along the line of sight, which change with viewing angle. DPR values $< 1$ are more common for a viewing angle of $0 \degree$, as there is no significant difference in the RM magnitude between the jets. The scatter in DPR values is much greater over time (up to a factor of 10 times greater values) than those seen by \citet{hardcastle_numerical_2014}; we compare multiple viewing angles to the radio lobes, whereas they consider their lobes at a viewing angle of $45\degree$ only. Our cluster simulation shows much greater scatter for all viewing angles since it is partially depolarised at $5.5$ GHz, in comparison to the group simulations which are not depolarised at that frequency.

\subsection{Comparison of Stokes parameters to previous work}
\label{section:stokes-comparison}

Our fractional polarisation maps are calculated from the Stokes parameters following Equation \ref{eqn:fracpolar}. We compare our Stokes Q and U parameters qualitatively with previous images rendered by \citet{hardcastle_numerical_2014} and \citet{english_numerical_2016} to understand the differences between this previous method for calculating the Stokes parameters and our new method which uses an evolving electron energy distribution. \citet{english_numerical_2016} use modified Stokes parameters that take into account aberration effects in relativistic fluids. Here, we follow \citet{english_numerical_2016} to calculate the Stokes parameters for our Lagrangian particles:

\begin{align}
\label{eqn:stokesHK14}
\begin{split}
    j_I &= p(B_x^2 + B_z^2)^{(\alpha - 1)/2} (B_x^2 + B_z^2) D^{3 + \alpha}, \\
    j_Q &= \mu_P p(B_x^2 + B_z^2)^{(\alpha - 1)/2} (B_x^2 - B_z^2) D^{3 + \alpha}, \\
    j_U &= \mu_P p(B_x^2 + B_z^2)^{(\alpha - 1)/2} (2B_xB_z) D^{3 + \alpha}. \\
\end{split}
\end{align}

\begin{figure*}
    \centering
    \includegraphics{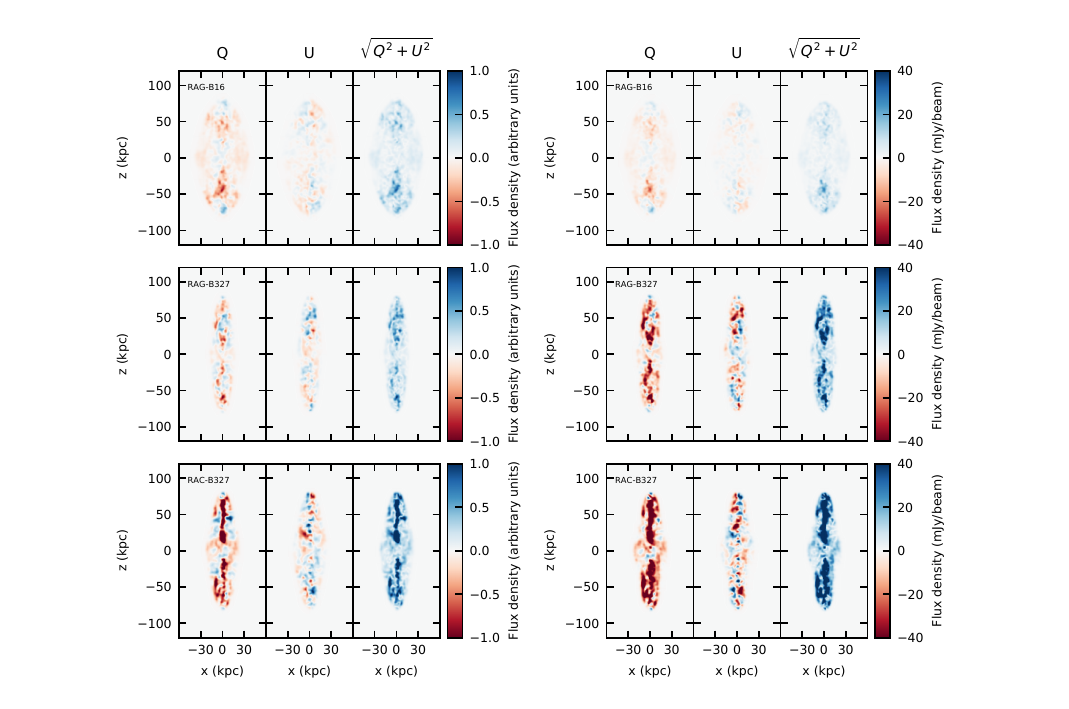}
    \caption[Comparison of Stokes' parameter calculations]{Stokes' parameters (Q, U, $\sqrt{Q^2+U^2}$) for simulations RAG-B16 (top row), RAG-B327 (middle row), and RAC-B327 (bottom row). Left: Stokes' parameters calculated using the method outlined in \citet{english_numerical_2016}. Right: Stokes' parameters calculated using the method described in this paper.}
    \label{fig:stokes-QU-noRM-comparison}
\end{figure*}

Here, the maximum fractional polarisation $\mu_P = (\alpha + 1)/(\alpha + 5/3)$, and the Doppler factor $D = 1/(\gamma (1 - \beta \cos\theta))$, where $\beta = v/c$ and $\theta$ is the angle between the line of sight and the velocity vector of the cell. We integrate the emissivities $j_I, j_Q,$ and $j_U$ for the particles along the line of sight on the grid, in the same manner as outlined in \citet{yates-jones_praise_2022} and plot Stokes Q, U, and $\sqrt{Q^2 +U^2}$ for each of our RMHD simulations in the left panel of Fig. \ref{fig:stokes-QU-noRM-comparison}. For consistency with our Stokes parameters, we scale the magnetic field by the volume filling factor as $B' = B / \sqrt{\rm trc}$ and multiply the Stokes emissivities by the volume filling factor from Equation \ref{eqn:volumefill_BRAiSE} for this calculation. We plot our Stokes Q and U parameters in the right panel of Fig. \ref{fig:stokes-QU-noRM-comparison}, where the parameters are calculated as detailed in Equation \ref{eqn:stokes_params}. For each of the calculation methods, we see the typical patchy effect from the complex magnetic field seen in Fig. \ref{fig:frac-polar}.  

Using the method described in this paper, we find that for simulation RAG-B16, there are minimal differences between the Stokes Q and U parameters between the two different calculation methods. However, for the higher jet magnetic field simulations, there are minor differences in both Stokes Q and U; in general, the images are broadly consistent with one another. For the \citet{english_numerical_2016} method, Stokes Q is in general more positive across the lobe, and the negative jet emission does not dominate as strongly as compared to our method. The Stokes U maps for the two methods have a similar distribution of positive and negative emission in the opposite sense to each other, due to a difference in sign between Equations \ref{eqn:chi0} and \ref{eqn:stokesHK14}; this sign difference does not affect the morphology of the structures. 

The differences seen for our higher jet magnetic field simulations are due to the different dynamics between these simulations and simulation RAG-B16; the jets with a higher magnetic field strength are more stable to deceleration. Simulations RAG-B327 and RAC-B327 both have disrupted but not fully decelerated jets and therefore relativistic effects will be important for the synthetic polarisation calculation. We conclude that the stability of the jet impacts calculation of the Stokes parameters using the \citet{english_numerical_2016} method due to the relativistic effects on the polarisation position angle. 

\section{Conclusions}
\label{section:conclusions_BRAiSE}

Using three-dimensional relativistic magnetohydrodynamic simulations of AGN jets and an extension for the synchrotron emission code PRAiSE \citep{yates-jones_praise_2022}, known as BRAiSE, we have demonstrated that the details of simulated radio synchrotron emission from AGN jets depends on whether the magnetic field or pressure is used to calculate it. This is due to the different distribution of magnetic energy density and pressure values throughout the lobe. This impacts both the spatial distribution of emission and the overall luminosity, thereby affecting the morphology and FR classification of the radio source. We also use the new BRAiSE method to study the polarisation properties of our simulated sources and the depolarising effect of Faraday rotation in both group and cluster environments.

High resolution surface brightness images produced using BRAiSE show clumps due to the complex magnetic field structure in the lobe. In general, BRAiSE produces sources with greater total luminosity due to the variation of magnetic field strength across the lobes. We simulate three sources; one that resembles a lobed FR-I source, and two that resemble FR-II sources. BRAiSE produces a diffuse hotspot region in our FR-II-like sources, since the magnetic energy at the jet head is spread out due to kink instabilities in the jet. BRAiSE also produces strong emission in a flaring point for these sources, which affects their classification using the FR index. We find that for PRAiSE, the FR index classifies each of our sources as an FR-II. However, for BRAiSE, these sources are more borderline. For our highly magnetised jets, BRAiSE predicts significantly lower FR indices than PRAiSE. We conclude that magnetic field information is required to interpret the FR morphology, and that our simulations show unique shock structures with the presence of both a flaring point and a hotspot that result in a bimodal surface brightness distribution. This unique shock structure indicates that we have not simulated true FR-II sources, and that our jet setup is missing the full set of physics required to stabilise the jet on large scales \citep[e.g. jet rotation;][]{bodo_linear_2016,bodo_linear_2019}.

Using BRAiSE, we have calculated Stokes I, Q, and U images for each of our RMHD simulations. For low and high resolutions and redshifts, we find patchy regions of high fractional polarisation. Over time, for each simulation, we find that the integrated fractional polarisation is lowest for sources viewed at an oblique angle. For our simulation with a lower jet magnetic field, this trend holds for the mean fractional polarisation. However, for our simulations with a higher jet magnetic field, the mean fractional polarisation is highest for this oblique viewing angle. This is due to the narrow morphology of these strongly magnetised sources.

Using Faraday rotation measures, we study the effect of depolarisation for our polarised sources: depolarisation is strongest at low frequencies, as expected. We confirm the Laing-Garrington effect for our group simulations, in that the upper radio lobes are in general less depolarised than the receding radio lobes. For our cluster simulation, we find that the RM values are higher and thus the Faraday depolarisation is a stronger effect than in the group environment for frequencies ranging between $0.15 - 5.5$ GHz, confirming the result from \citet{jerrim_faraday_2024}.

The method we use to calculate the Stokes Q and U is different to a previous method used by \citet{huarte-espinosa_3d_2011}, \citet{hardcastle_numerical_2014}, and \citet{english_numerical_2016}. We qualitatively compare the differences between the two methods for our simulations, finding that our simulations with a higher jet magnetic field strength display minor differences in their Stokes Q and U maps, however, the simulation with a low magnetic field strength is almost identical. This demonstrates that the relativistic aberration effects should be accounted for in simulations of relativistic radio lobes.

Characterisation of magnetic field structure and synchrotron emissivity enabled by our RMHD simulations and analysis permits investigations of several other aspects of radio galaxy physics. We will compare and contrast more of the Stokes I spectral properties in our next paper. This and future work will be used in the CosmoDRAGoN project \citep{yates-jones_cosmodragon_2023} to study the synthetic emission of RMHD jets in realistic, magnetised, cluster environments. This project will provide insights on the link between observed radio AGN sources and the underlying physical mechanisms of AGN jet feedback in galaxy clusters.

\section{Acknowledgements}

LJ thanks the University of Tasmania for an Australian Government Research Training Program (RTP) Scholarship. SS, CP, and PYJ acknowledge funding by the Australian Research Council via grant DP240102970. This research was carried out using the high-performance computing clusters provided by Digital Research Services, IT Services at the University of Tasmania. This work has made use of data from The Three Hundred collaboration (https://www.the300-project.org) which benefits from financial support of the European Union’s Horizon 2020 Research and Innovation programme under the Marie Skłodowskaw-Curie grant agreement number 734374, that is the LACEGAL project. The Three Hundred simulations used in this paper have been performed in the MareNostrum Supercomputer at the Barcelona Supercomputing Center, thanks to CPU time granted by the Red Española de Supercomputación. We thank \citet{knapen_how_2022} for their guide to writing astronomy papers. We acknowledge the support of the developers providing the Python packages used in this paper: Astropy \citep{astropy_collaboration_astropy_2022}, JupyterLab \citep{kluyver_jupyter_2016}, Matplotlib \citep{hunter_matplotlib_2007}, NumPy \citep{harris_array_2020}, and SciPy \citep{virtanen_scipy_2020}.

\paragraph{Data Availability Statement}

The data underlying this article will be shared on reasonable request to the corresponding author.


\bibliography{references.bib}

\appendix

\section{Plasma composition}
\label{section:appendix}

The fluid tracer in PLUTO describes the fraction (by mass) of radiating fluid within a grid cell in the simulation compared to the non-radiating (thermal) fluid. The passive Lagrangian particles record the history of the fluid values (including the tracer value and vector magnetic field) as they are advected with the fluid on the simulation grid. To calculate the synchrotron emission from these particles, we are interested in the fraction of this particle that is radiating.

Therefore, the synchrotron emissivity weighting depends on the assumed mass ratio between the radiating and thermal plasmas. For an electron-proton (`normal') plasma, this ratio is unity (as $m_r = m_{th} = \mu m_p$), the volume filling factor simplifies to $f = {\rm trc}$, and the emissivity of each particle is simply weighted by its tracer value. This is the approach implemented in \citet{yates-jones_praise_2022}.

We also consider the case of an electron-positron (`pair') plasma, where the volume filling factor is $\sim 1$ for tracer values $\gg \frac{m_r}{m_{\rm th}} = 9 \times 10^{-4}$, and will decrease quickly for tracer values $\ll 9 \times 10^{-4}$. For a tracer value $\simeq 9 \times 10^{-4}$, the volume filling factor is $\simeq 0.5$. Effectively, this is the same as removing the emission from all low tracer particles.

\begin{figure*}
    \centering
    \includegraphics{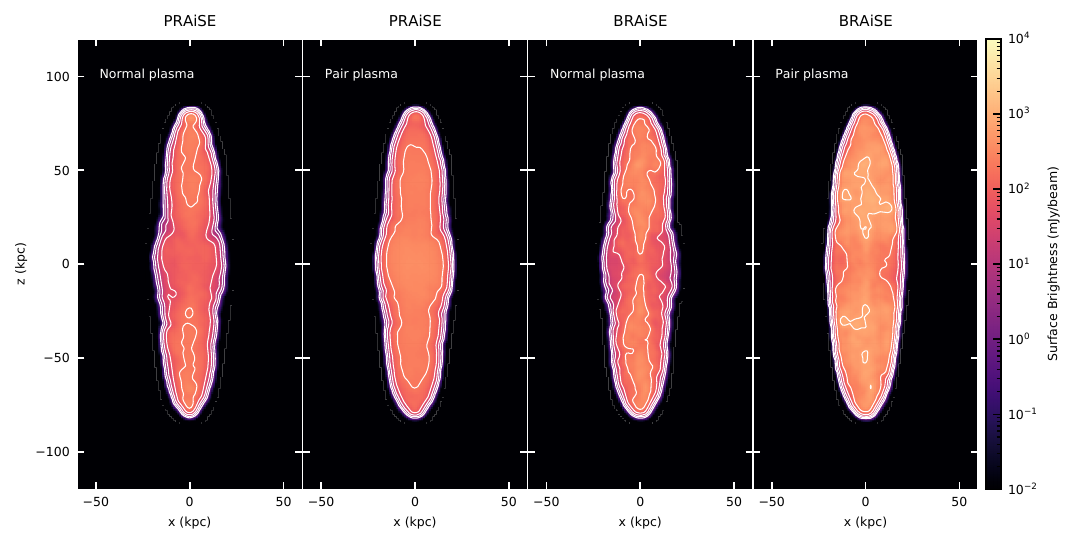}
    \caption[Plasma comparison plots for simulation RAG-B327]{PRAiSE (left two panels) and BRAiSE (right two panels) surface brightnesses (Stokes' I) for the RAG-B327 simulation at $8$ Myr, $z = 0.05$, and $0.15$ GHz. Contours are at $1, 3.72, 13.9, 51.8, 193, 720, 2683,$ and $10000$ mJy/beam. Two plasma compositions are considered: `normal' (mass ratio $ = 1$) and `pair' (mass ratio $= 9 \times 10^{-4}$).}
    \label{fig:sb-plasmacomp}
\end{figure*}

We plot the surface brightness for weighting by the two different plasma compositions in Figure \ref{fig:sb-plasmacomp}. We find that assuming a pair plasma results in a radio source with structures that are less characteristic of classical double (FR-II) sources. This effect is more pronounced for the PRAiSE method, where there is a significant amount of emission in the equatorial region and the hotspots at the ends of the lobes are not clearly defined. Similarly for the BRAiSE method, the hotspot emission is much less clearly defined, however, the emission in the equatorial region is not as significant. Therefore, we use the `normal' plasma weighting for the synthetic radio emission results presented in this paper, consistent with \citet{yates-jones_praise_2022}.

\end{document}